\documentclass[aps,prd,a4paper,superscriptaddress,nofootinbib,showpacs,showkeys,amsfonts,amssymb,amsmath]{revtex4}
\usepackage{amssymb,latexsym}
\usepackage{amsmath, amsthm}
\usepackage{amscd}
\usepackage{times}
\usepackage{epsfig}
\usepackage{psfrag}
\usepackage{graphicx}
\usepackage{overpic}
\usepackage[english]{babel}

\newcommand{\be}{\begin{equation}}
\newcommand{\ee}{\end{equation}}
\newcommand{\bea}{\begin{eqnarray}}
\newcommand{\eea}{\end{eqnarray}}

\begin{document}

\title[]{Classical collapse to black holes and quantum bounces: A review}

\author{Daniele Malafarina}
\email{daniele.malafarina@nu.edu.kz}
\affiliation{Department of Physics,
Nazarbayev University, 53 Kabanbay Batyr avenue, 010000 Astana, Kazakhstan}

\begin{abstract}
In the last four decades different programs have been carried out aiming at understanding the final fate of gravitational collapse of massive bodies once some prescriptions for the behaviour of gravity in the strong field regime are provided. The general picture arising from most of these scenarios is that the classical singularity at the end of collapse is replaced by a bounce. The most striking consequence of the bounce is that the black hole horizon may live for only a finite time. 
The possible implications for astrophysics are important since, if these models capture the essence of the collapse of a massive star, an observable signature of quantum gravity may be hiding in astrophysical phenomena. One intriguing idea that is implied by these models is the possible existence of exotic compact objects, of high density and finite size, that may not be covered by an horizon.
The present article outlines the main features of these collapse models and some of the most relevant open problems. The aim is to provide a comprehensive (as much as possible) overview of the current status of the field from the point of view of astrophysics.
As a little extra, a new toy model for collapse leading to the formation of a quasi static compact object is presented.
\end{abstract}


\maketitle

\section{Introduction}\label{intro}

Our present understanding of the universe and its evolution implies the existence of black holes, bodies whose masses are packed in such small volumes that not even light can escape. 
We have experimental evidence that such objects do exist from observations of x-ray binaries, which suggests the existence of stellar mass black holes in binary systems, and from the spectral properties of quasars and active galactic nuclei, which suggest that super-massive black holes dwell at the center of most galaxies.

From a theoretical point of view, black holes are a direct consequence of the fact that we must use General Relativity (GR) to describe the late stages of gravitational collapse. For collapsing matter sources satisfying standard energy conditions, Einstein's field equations imply that eventually collapse must lead to the formation of trapped surfaces and a singularity
\cite{sing1}, \cite{sing2}, \cite{HE}.
The generic existence of singularities in solutions of Einstein's field equations is a troublesome issue for classical GR as their presence signals a regime where predictability breaks down
\cite{hawk}
and the theory doesn't hold. 
The most conservative view on singularities is that they are a consequence of the application of the theory in a regime where quantum effects become important and thus they should not appear in a full theory of quantum gravity. This point of view is usually traced back to Wheeler (see for example 
\cite{wheeler}
for an historical overview).

Our theoretical understanding of how black holes form is rooted in the simplest toy model for spherical collapse that was developed by Oppenheimer and Snyder
\cite{OS} 
and independently by Datt \cite{datt} 
in 1939 (from now on referred to as the OSD model). 
On the one hand, we know that GR works very well in the weak field and we are confident that models of collapse such as OSD are accurate to describe the fundamental features of collapse scenarios far away from the singularity.
On the other hand, the behavior of gravity in the strong field is not well understood and even though we do know that GR requires modifications in the regime where the gravitational field becomes strong over very small scales, we still do not know what kind of form such modifications should take.
Black hole horizons somehow stand at the crossroad between these two situations. For example, the event horizon for a Schwarzschild black hole sits comfortably far from the strong gravity region but its description is strongly linked to the existence of the singularity and modifications to our classical models in the vicinity of the singularity bear important consequences for the behavior of the horizon itself.

The question of how gravity behaves in the strong field is also tightly connected to the question `What happens to matter when it is compressed to volumes small enough that the classical description fails?' 
Therefore one could expect that both sides of Einstein's equations (the geometrical side containing Einstein's tensor $G_{\mu\nu}$ and the matter side containing the energy-momentum tensor $T_{\mu\nu}$) will need to be modified in the strong field regime.

The investigation of simple analytical toy models in quantum gravity is motivated by the idea that these scenarios could provide valuable information about the general features that a full theory of quantum gravity should posses.
The general view that is taking shape in the last couple of decades is that quantum effects will generate repulsive pressures that are sufficient to counteract the gravitational attraction, thus avoiding the singularity at the end of collapse. The first, immediate consequence, is that it should be impossible to form a Schwarzschild (or Kerr) black hole within a full theory of quantum gravity.
Then different possibilities may arise depending on the assumptions made in the model. For example, collapse may stop before the formation of the horizon, leaving behind an exotic compact object. Alternatively collapse may lead to the formation of the horizon and eventually halt at much smaller scales. In this last, case collapsing matter would eventually bounce thus leading to a phase of re-expansion following collapse. The re-expanding phase may not affect the geometry in the exterior, thus leaving an object that looks like a black hole for distant observers, or it may trigger a transition of the black hole horizon to a white hole horizon.

In the present article we focus on modifications to dynamical scenarios where the black hole forms from regular initial data. However, it should be noted that the issue of how the static Schwarzschild black hole solution can be altered by the introduction of quantum effects has been addressed by several authors in different frameworks.
For example, solutions within classical GR describe what are now called `regular black holes'
(see for example \cite{bardeen} and \cite{hayward} for the earliest results, or \cite{fro} and \cite{boj4} for more recent discussions). These are modifications of the Schwarzschild solution that imply a minimum length scale and a non vanishing energy-momentum tensor. Rotating regular black holes were considered for example in
\cite{BM}, 
their properties as candidates for astrophysical objects were studied for example in
\cite{bobo1} and \cite{bobo2},
while the extension to the case with non vanishing cosmological constant was studied in
\cite{saa}.
Other solutions that modify the Schwarzschild black hole can be  obtained via the introduction of quantum effects. These are vacuum solutions within a quantum gravitational description of the black hole space-time. For example, approaches based on Loop Quantum Gravity (LQG) have been studied in \cite{ash3}, \cite{boj2}, \cite{pullin} and \cite{modesto}.
Other approaches have been also considered. 
For example, an improved Schwarzschild solution based on renormalization group and running gravitational constant was described in 
\cite{Bonanno},
while a quantization approach based on canonical formalism was presented in
\cite{kuns1} 
and a discussion of the meaning of the gravitational Schwarzschild radius within a quantum theory was presented in
\cite{casadio2}. 
To this aim a quantum mechanical operator acting on the `horizon wave-function' was introduced.
All these models show similar features, like, for example, the appearance of an inner horizon inside the event horizon (see for example 
\cite{torres} for a study on the properties of the inner horizon of the solution presented in \cite{Bonanno} and \cite{torres-new} for a more recent study with cosmological constant) or the possible existence of a massive, dense compact object of finite size smaller than the horizon. However the most important insight coming from the study of quantum corrected black holes is the realization that quantum effects may modify the geometry at large scales thus having implications for the description of the horizon of the black hole.

The focus of the present article is on existing dynamical models of collapse, and on how the introduction of `quantum' effects may alter the standard picture of black hole formation. 
We will review the current status of our understanding of gravitational collapse when some kind of repulsive effects (that can be interpreted as quantum corrections) are incorporated as the gravitational field becomes strong.
In dynamical models the main mechanism that leads to the avoidance of the singularity is when collapse halts at a finite radius, possibly producing a bounce for the infalling matter. 
As mentioned before, the cloud may halt at a radius larger than the Schwarzschild radius, thus never forming a black hole. In this case it may produce a compact object or it may completely evaporate (see for example, \cite{grava1} for an example of compact object, \cite{rad2}, \cite{rad3} and \cite{mersini} for the effects of black hole evaporation on gravitational collapse and \cite{yuki1}, \cite{yuki2} for examples of evaporation including 4D Weyl anomaly). 
Alternatively the cloud may reach a minimum scale and re-expand. If the the repulsive effects close to the bounce influence the geometry in the vacuum exterior then the black hole effectively turns into a white hole (see for example \cite{garay2}, \cite{rov1} and \cite{bar3}). 
On the other hand, if repulsive effects are confined to a small neighborhood of the center then external observers would effectively see a black hole (see for example \cite{frolov}).
Our aim is to point out the main outstanding unresolved issues in theoretical models and how such models might be tested against observations of energetic phenomena in the universe in the near future.


The first proposed model of this kind involved the collapse of a thin light-like shell in a quantum gravity scenario in which the effective Lagrangian was described by the Einstein's Lagrangian plus the leading higher order terms that describe gravity over short distances 
\cite{frolov}, \cite{frolov2}.
There it was shown for the first time that quantum gravitational effects can avoid the occurrence of singularities and, as a consequence, shorten the life of the black hole horizon.

In more recent times the renewed interest for quantum corrections to classical singularities in collapse models was sparked by Loop Quantum Cosmology (LQC). Several authors, based on considerations coming from LQG, showed how quantum corrections near the big bang singularity can remove the initial singularity and produce a bouncing universe
(see for example 
\cite{ash}, \cite{ash2} and \cite{boj}).
Since the simplest relativistic toy models for collapse are essentially the time reversal of big bang models, the same formalism derived from LQC can in principle be applied to collapse
(see for example \cite{boj3}). 

However, the collapse of a massive star is very different from the evolution of the universe. Above all, the most notable difference that one has to deal with, when considering collapse, is the matching to an exterior manifold describing the space-time around the collapsing object. In the OSD model, an horizon develops in the exterior once the infalling matter crosses the Schwarzschild radius. 
How the bounce will affect the horizon in the vacuum exterior is presently still not entirely clear. If quantum effects are confined inside the collapsing matter then the light-cone structure of the space-time must undergo a discontinuous transition from the interior to the exterior. On the other hand if quantum effects propagate in the exterior then the black hole geometry must be altered.

As mentioned above, most dynamical models with `quantum' corrections studied to date result in a bouncing scenario. These models suggest the possibility of the formation of exotic compact remnants as leftovers from collapse and recently there has been a lot of interest in the possible phenomenology of such remnants.
One could ask whether a regular black hole can form through a dynamical process
(see for example \cite{bam-mod}), 
or what kind of properties such remnants would have
(see for example \cite{vidotto}).
The idea that compact objects other than neutron stars can form from collapse has been around for a long time. For example, exotic compact objects were proposed by Hawking already in 1971
\cite{hawking}.
Just like the electron degeneracy pressure can stabilize collapse leading to a white dwarf and the neutron degeneracy pressure can stabilize it to produce a neutron star, it seems reasonable to suppose that a further island of stability may exist at densities higher than neutron star's cores for a yet unknown state of matter. 
Along these lines several kinds of exotic compact objects have been proposed through the years. For example objects like gravastars (`gravitational vacuum stars', obtained from a phase transition of quantum vacuum near the location of the horizon) 
(see \cite{grava1} and \cite{grava2})
and black stars 
(see \cite{bar1})
have a radius slightly larger than the Schwarzschild radius. 
On the other hand theoretical objects like
quark stars
(see \cite{quark1} and \cite{quark2})
and boson stars
(see \cite{boson1} and \cite{boson2})
have a larger boundary (larger than the photon sphere for black holes).
The properties of these proposed objects have been studied in detail.
Also, arguments for the existence of compact remnants left over after black hole evaporation were proposed in connection with the information loss problem
(see for example \cite{gid1}, \cite{chen} and \cite{lochan} and references therein). In these scenarios, a compact object of Planck scale may be the residue of the complete evaporation of the black hole via Hawking radiation.
The question of which kind of dynamical process may lead to the formation of such objects remains open and it is closely connected to how matter and gravity behave at extremely high densities. 

The paper is organized as follows: In section \ref{class} we review the OSD model for dust collapse and Einstein's equations for the collapse of homogeneous dust and perfect fluids (the reader familiar with such topics can jump directly to the next section). 
Section \ref{semiclass} is devoted to a review of semi-classical corrections to collapse and their consequences for the OSD model and black hole formation in general. 
In section \ref{open} we outline the main open questions related to such modified collapse models. 
In section \ref{pheno} we explore the phenomenological consequences that semi-classical collapse bears for astrophysical black holes and 
we investigate the possibility that exotic compact objects and remnants may occur as leftover from collapse. In section \ref{pheno} we also introduce a new dynamical toy model leading to one of such hypothetical remnants (which we call a dark energy star). 
Finally section \ref{discussion} is devoted to a brief discussion of the present and future status of the field.

Bullet points are used throughout the sections in order to highlight the separation of each topic from the next.
Finally, throughout the paper we will make use of geometrical units by setting $G=c=1$ and for simplicity, we will absorb the factor $8\pi$ in Einstein's equations into the definition of the energy momentum tensor.


\section{Classical collapse...}\label{class}

In order to set the stage we will briefly review here the formalism for gravitational collapse of matter fluids in GR.
For the interior of the collapsing cloud we consider a spherically symmetric space-time described in co-moving coordinates by the metric
\be \label{interior}
ds^2=-e^{2\nu(r,t)}dt^2+e^{2\Phi(r,t)}dr^2+R(r,t)^2d\Omega^2 \; ,
\ee 
where $d\Omega^2$ is the line element on the unit sphere and $\nu(r,t)$, $\mu(r,t)$ and $R(r,t)$ are the metric functions to be determined from Einstein's equations.
The energy momentum tensor is that of a gravitating fluid given by $T_{\mu\nu}={\rm diag}(\rho(r,t),p_r(r,t),p_\theta(r,t),p_\theta(r,t))$ for an anisotropic source. In the following we will concentrate on perfect fluids with $p_r=p_\theta=p$.
The metric \eqref{interior} is continuously matched to a know exterior (such as Schwarzschild or Vaidya) through the collapsing boundary $R_b(t)=R(r_b(t),t)$
(see \cite{matching1}-\cite{matching4} 
for details).
Then Einstein's equations can be written in the form
\bea \label{rho}
 \rho&=& \frac{F'}{R^2R'} \; , \\ \label{p} 
 p_r&=&-\frac{\dot{F}}{R^2\dot{R}} \; , \\ \label{nu} 
 \nu'&=&2\frac{p_\theta-p_r}{\rho+p_r}\frac{R'}{R}-\frac{p_r'}{\rho+p_r} \; , \\ \label{G}
 \dot{R}'&=&\frac{1}{2}\left(R'\frac{\dot{G}}{G}+\dot{R}\frac{H'}{H}\right) \; ,
\eea
where primed quantities denote derivative with respect to the co-moving radius $r$ and dotted quantities denote derivatives with respect to the co-moving time $t$. The function $G(r,t)$ is related to the initial velocity of the particles in the cloud and $G$ and $H$ are defined as
\be 
G=e^{-2\Phi}R'^2, \; \; H=e^{-2\nu}\dot{R}^2 \; .
\ee
The function $F(r,t)$ is called the Misner-Sharp mass 
of the system 
\cite{misner}, 
it describes the amount of matter contained within the co-moving radius $r$ at the co-moving time $t$ and it takes the form
\be \label{misner}
F=R(1-G+H) \; .
\ee 
The system consists of five equations in seven unknown functions, therefore in order to close it one has to specify two equations of state, for the radial pressure and the tangential pressure. For example, by setting $p_r=p_\theta=0$ one obtains a cloud of non interacting particles (i.e. `dust'), homogeneous perfect fluids are given by $p_r=p_\theta=p(t)$, in this case a linear equation of state $p=\lambda\rho$ is often considered. More realistic mater fluids can be described by a polytropic equation of state of the kind $p_r=p_\theta=K\rho^\gamma$ like the one used to describe equilibrium configurations
\cite{tooper}. 
Other, more exotic, possibilities may be considered as well.
Equation \eqref{misner} can be rewritten as
\be \label{motion}
\dot{R}=\pm e^\nu\sqrt{\frac{F}{R}+G-1} \; ,
\ee 
and treated as the equation of motion describing the trajectory of each collapsing shell of matter. Note that in order to describe collapse one has to take the minus sign in the equation above.
Trapped surfaces develop in the interior when
\be \label{horizon}
1-\frac{F}{R}=G-e^{-2\nu}\dot{R}^2=0 \; ,
\ee
which implicitly defines the curve $t_{\rm ah}(r)$ describing the co-moving time at which the shell $r$ becomes trapped.

It is generally useful to make use of a gauge degree of freedom and rewrite the equation of motion in terms an adimensional scale factor $a(r,t)$ (for details see for example
\cite{review}). 
This in turn allows for the introduction of a similar scaling for the Misner-Sharp mass $F$ and the velocity profile $G$, as
\begin{itemize}
 \item Scale factor $a(r,t)$ given by: $R=ra$.
 \item Mass function $M(r,t)$ given by: $F=r^3M$.
 \item Velocity profile $b(r,t)$ given by: $G=1+r^2b$.
\end{itemize}
Then, the equation of motion \eqref{motion} can be rewritten as
\be\label{motion2}
\dot{a}=-\sqrt{\frac{M}{a}+b} \; .
\ee



For collapse of realistic fluids one usually considers a matter model satisfying the weak, strong or dominant energy conditions. The least stringent of such requirements is given by the weak energy conditions that translate to $\rho>0$, $\rho+p_r>0$ and $\rho+p_\theta>0$.
Of course, all types of matter observed in the universe today satisfy the classical energy conditions mentioned above. `Exotic' matter sources have been conjectured but not observed so far. Nevertheless it is reasonable to suppose that in the ultra-dense regime where quantum gravitational effects become important matter will satisfy some quantum version of the energy conditions.
For example in \cite{visser} it was shown that non linear energy conditions are more suitable to describe matter in a regime transitioning from a classical to a quantum state.

At a classical level it is well known that in a globally hyperbolic space-time, any matter field that satisfies energy conditions and is compact enough to form a trapped surface must inevitably form a singularity as well (for a complete historical review of the singularity theorems see \cite{senovilla-garfinkle}).
Therefore, within the framework of classical GR, the removal of the singularities at the end of collapse is closely related to the violation of at least one of the assumptions of the singularity theorems.
From the above considerations then it is easy to conclude that the most natural way to avoid such singularities is to violate the energy conditions at some stage during collapse. This approach was first followed in 
\cite{berg}.
Violation of energy conditions implies repulsive effects and these effects are responsible for halting collapse and avoidance of the singularity.
There are essentially two ways in which the classical energy conditions may be violated in the strong field regime: We can assume that we can still use GR to describe the evolution of the collapsing cloud but the matter fields, within some quantum theory of gravity, are such that they violate the energy conditions. Or we can assume that matter satisfies the energy conditions and repulsive effects arise from quantum modifications to GR. As we will see below, such modifications to GR may be treated as an effective matter source to be put on the right-hand side of Einstein's equations. In this way the problem is reduced to that of determining the evolution of an effective, non physical, source within classical GR.

\subsection{Dust, homogeneous fluids and null shells}

\begin{figure}[tt]
\centering
\begin{minipage}{.45\textwidth}
\centering
\includegraphics[scale=0.3]{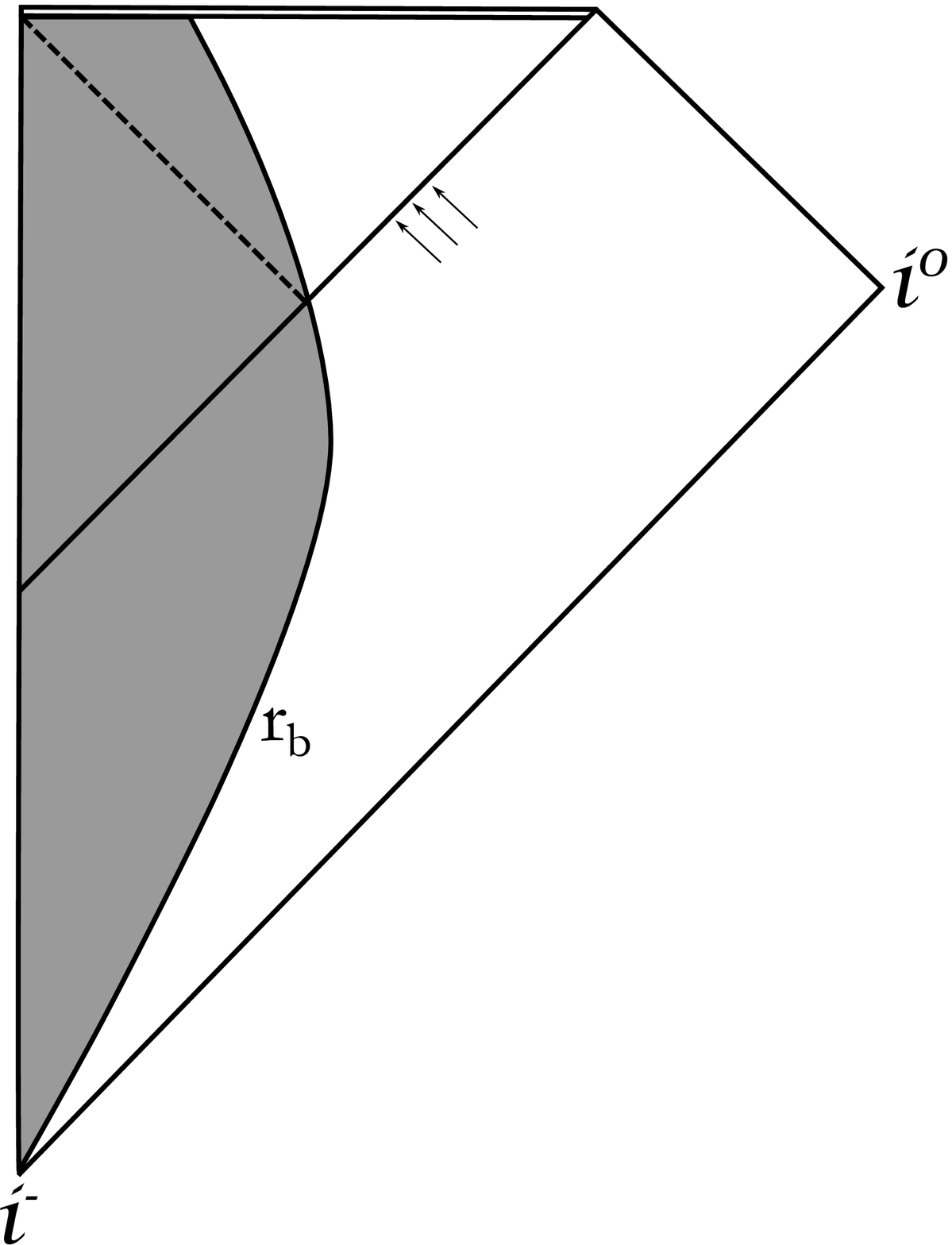}
\put(-120,90){I}
\put(-65,120){II}
\end{minipage}
\hfill
\begin{minipage}{.45\textwidth}
\centering
\includegraphics[scale=0.3]{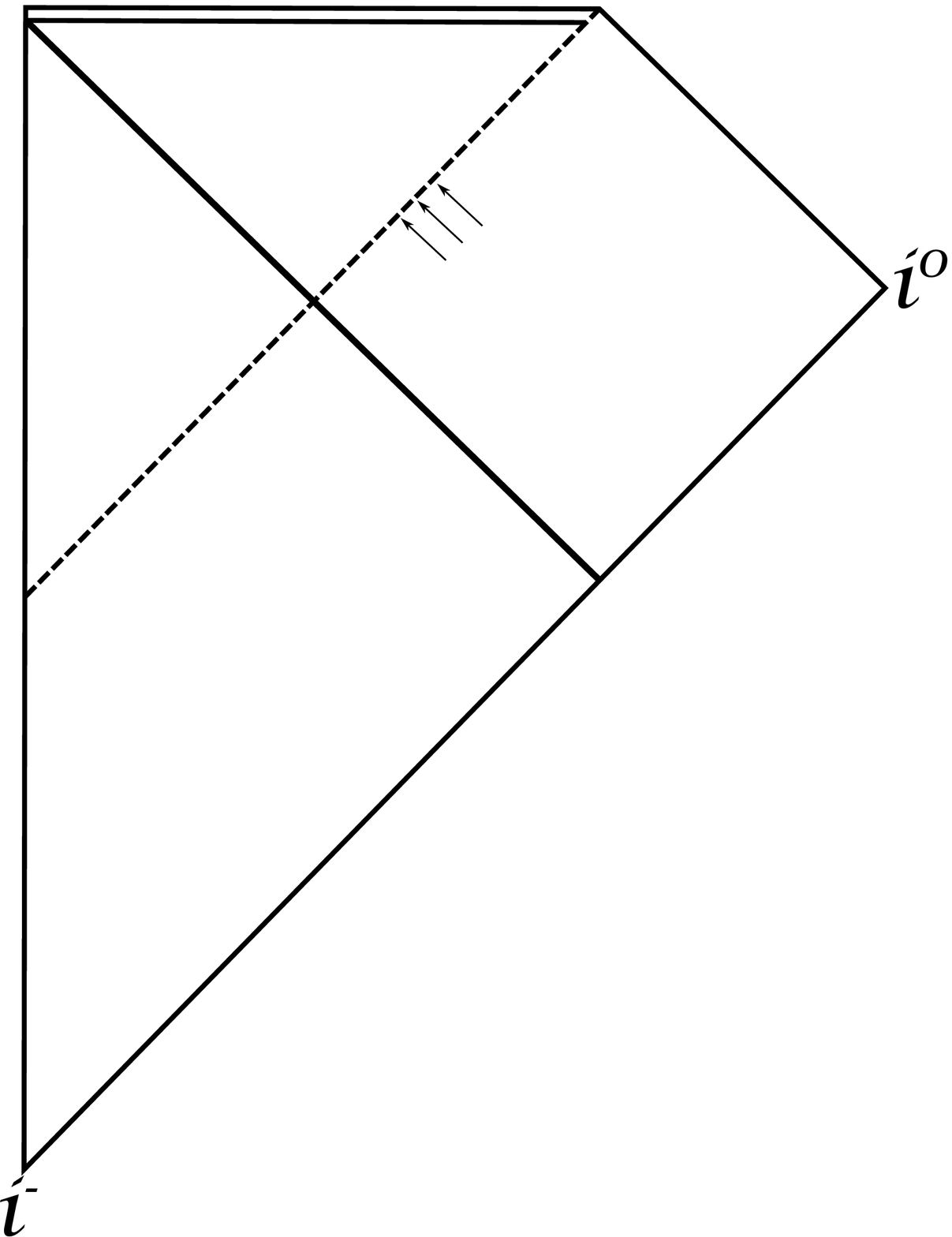}
\put(-100,94){I}
\put(-50,140){II}
\end{minipage}
\caption{Penrose diagrams for collapse with the formation of a singularity (double solid line). 
Left panel: Collapse of a spherical dust cloud. The solid curved line $r_b$ represents the boundary of the cloud. The interior region I, is described by pressureless particles (grey area), the exterior region II, is described by the Schwarzschild vacuum space-time. As the boundary passes the Schwarzschild radius the trapped region develops. In the exterior the event horizon (solid diagonal line) forms at $r=2M$, while in the interior the apparent horizon (dashed line) moves inwards from the boundary towards $r=0$.
Right panel: Collapse of a thin null shell (thick line) separating a Minkowski interior, region I, from a Schwarzschild exterior, region II. The event horizon (dashed line) meets the collapsing shell at the Schwarzschild radius. The null fluid focuses at the center forming a singularity at $r=0$.}
\label{fig1}
\end{figure}

\textbullet\ The simplest matter model that one can consider to study collapse is the OSD model describing homogeneous dust (i.e. non interacting particles). In this case $\rho=\rho(t)$ and $p_r=p_\theta=0$.  From equation \eqref{p} it follows that the Misner-Sharp mass must be $F(r)=r^3M_0$, with $M_0$ an arbitrary constant. Then Einstein's equations simplify dramatically and it's easy to see that we can set $\nu=0$ and $G=1+r^2b_0$ with $b_0$ an integration constant. In the marginally bound case given by $b_0=0$ (where ideally particles have zero initial velocity at spatial infinity) solving the set of Einstein's equations reduces to solving the equation of motion \eqref{motion2} which 
is readily integrated with the initial condition $a(0)=1$ to give
\be 
a(t)=\left(1-\frac{3}{2}\sqrt{M_0}t\right)^\frac{2}{3} \; .
\ee 
Then it is immediately seen that the singularity forms at the time $t_s=2/(3\sqrt{M_0})$ when $a=0$. From equation \eqref{rho} we see that the energy density is $\rho(t)=3M_0/a^3$ and it is diverging as $t$ goes to $t_s$. The event horizon in the exterior region develops at the time $t=t_s-2r_b^3M_0/3<t_s$, at the same time the apparent horizon develops in the interior. The apparent horizon moves inwards from the boundary to reach the center at the time of formation of the singularity.
The whole scenario is summarized in the well known Penrose diagram in the left panel of figure \ref{fig1}.

\textbullet\ A similar situation occurs when one considers collapse of an homogeneous perfect fluid with linear equation of state of the type $p=\lambda\rho$. Then the dominant energy condition implies that $\lambda\in[-1,1]$. Einstein's equations simplify in a similar way to the dust case and we now obtain a variable Misner-Sharp mass $M(t)=M_0/a^{3\lambda}$, which implies an inflow or outflow of matter through the boundary (matching should then be performed with a Vaidya exterior).
The equation of motion for marginally bound collapse is also easily integrated to give 
\be 
a(t)=\left(1-\frac{3(\lambda+1)}{2}\sqrt{M_0}t\right)^{\frac{2}{3(\lambda+1)}} \; .
\ee
We can see that collapse proceeds in a way qualitatively similar to the dust case (note that $\ddot{a}$ changes sign when $\lambda$ becomes smaller than $-1/3$).
For dust and homogeneous fluids the condition for the formation of trapped surfaces can be rewritten as
\be \label{ah}
1-\frac{r^2M}{a}=0 \; ,
\ee
which gives implicitly the curve $t_{\rm ah}(r)$ describing the co-moving time at which the co-moving shell $r$ becomes trapped.

\textbullet\ The situation for collapse of a null shell is slightly different as one has to use the Vaidya space-time 
\cite{vaidya}. 
In the case of collapse of a thin shell the space-time can then be divided into three regions: A Minkowski interior that is separated from a Schwarzschild exterior by the collapsing thin null shell
(see for example \cite{joshi-book}).
The metric can then be written in advanced Eddington-Finklestein coordinates as
\be 
ds^2=-\left(1-\frac{2M}{r}\Theta(v)\right)dv^2+2dvdr+r^2d\Omega^2 \; ,
\ee 
where $\Theta(v)$ is the Heaviside function separating the interior from the exterior. Again it is easy to see that the event horizon in the exterior forms as the null shell passes the radius $r=2M$ and eventually the singularity forms when the collapsing shell reaches the center. The whole scenario can be summarized in the Penrose diagram in the right panel of figure \ref{fig1}.
Collapse of null fluids can also be considered. In this case the Vaidya null dust space-time is not confined to a shell. Einstein's equations for collapse can be written in a similar manner as above and collapse may result in the formation of a black hole or a naked singularity depending on the behaviour of the mass function $M(v)$ 
(see \cite{joshi-book} for details).

These are simple toy models, which are not very realistic when we think about the collapse of a massive star. However, it is generally accepted that they encode all the relevant features to describe the formation of a black hole. Then, if we agree that more realistic models would be qualitatively similar, the main drawback of these models is the occurrence of the singularity, which signals their inability to capture what happens when the gravitational field becomes very strong over very short distances.
On the other hand one could question how the general picture of collapse given above would change if more realistic assumptions were considered.

\subsection{Toy models vs realistic models}

The collapse models described above have the obvious advantage of being simple. Equations can often be solved analytically and the global structure of the space-time can be described. However, it is also obvious that marginally bound collapse of a non rotating dust ball is hardly a realistic model to describe the collapse of a star.

Before moving on to quantum corrections in the strong field, in this section we shall outline the main features that one should address in order to describe a more realistic collapse scenario within classical GR. Typically toy models consider marginally bound collapse of homogeneous, isotropic, spherically symmetric fluids, with simple equations of state.
All of the above assumptions are considered in order to simplify the equations. However one could argue on the physical validity of the resulting models and whether the dynamical evolution of collapse would remain unchanged (at least qualitatively) if more realistic assumptions were made.

\textbullet Initial velocity: Marginally bound collapse means that the initial velocity profile corresponds to a configuration where particles have zero initial velocity at spatial infinity. This corresponds to taking the integration function $b$ coming from equation \eqref{G} to be zero.
From a geometrical point of view, for homogeneous models, this is equivalent to requiring that the interior geometry of the collapsing sphere be flat. However, realistic collapse is expected to start with zero initial velocity from a finite radius. Therefore assuming marginally bound collapse is an oversimplification that may hinder the physical validity of the models. For example, in
\cite{barcelo-uni}
the authors considered semi-classical models for collapse which starts from rest from a finite radius by using Painlev\'e-Gullstrand coordinates\footnote{The choice of Painlev\'e-Gullstrand coordinates is particularly well suited for describing collapse that starts from rest at a finite radius (see \cite{zip1} for a detailed discussion).}. 

\textbullet\ Homogeneity and isotropy: The choice of density and isotropic pressures depending only on $t$ helps to simplify the equations. However realistic fluids are expected to have density and pressure gradients. The main consequence of considering inhomogeneous clouds is that the boundary radius of the object can not be specified at will, like in the dust case, but must be determined by the condition of vanishing of the pressure. Also considering anisotropic fluids implies that the regime where the classical description fails would depend on the direction, thus complicating considerably the model.

\textbullet\ Equation of state: Most of the collapse models studied in the literature consider oversimplified models for matter. The OSD model is described by homogeneous dust, other scenarios typically neglect pressure gradients, heat conduction, viscosity and so on. Furthermore local energy conservation for such adiabatic perfect fluids implies that the entropy is constant in any given co-moving volume of the fluid.
If one wishes to give a more realistic description of collapse then a better matter model should be prescribed, where dissipative effects are present. Once a non trivial equation of state is introduced one would need to consider the fluid's hydrodynamics and its thermodynamical properties. In this case the boundary surface $r_b$ needs to be described by a surface layer that transports energy and momentum between the interior and the exterior, thus implying that the exterior can not be given by the Schwarzschild solution (typically a radiating Vaidya metric can be used
\cite{matching3}). In this case one needs to specify a varying boundary radius $r_b(t)$.
A first step towards an equation of state for stars that better describes realistic matter could be that of a polytropic fluid of the kind $p=K\rho^\gamma$, with $\gamma=4/3$
\cite{tooper}. 
Then, once the pressure and density at the center are given, the boundary of the object would be defined by the imposition of vanishing of the pressure, $p(r_b,t)=0$. 
It must be noted that any equation of state that holds in the weak field is unlikely to remain unchanged during the last stages of collapse. Therefore a prescription for the transition of matter towards the high density phase should be prescribed as well.
Unfortunately, as the equations describing the fluid in the interior become more complicated (involving hydrodynamics, turbolence, shock waves and so on) the hope to find an analytical solution must be abandoned and one has to resort to numerical simulations.

\textbullet\ Spherical symmetry: Given the lack of a viable interior for the Kerr metric it is not surprising that analytical dynamical models of collapse with rotation have not been explored in much detail. Of course one could study collapse scenarios of rotating matter fields without worrying about the matching with an exterior solution. However in order to be physically valid such models would still need to describe well behaved matter sources (typically sources that satisfy energy conditions, have well behaved density profile and are non singular at the initial time). In this respect models with slow rotation have been considered in the past
(see for example \cite{rotation1} and \cite{rotation2}).
The exterior field is usually taken as a vacuum solution with slow rotation, such as the Hartle-Thorne metric 
\cite{HT}.
This is a solution of Einstein's field equations in vacuum that describes the exterior of a slowly and rigidly rotating, stationary and axially symmetric body (up to an expansion in the angular momentum).
For years scientists wondered whether the final singularity appearing in OSD collapse was an artifact due to the choice of spherical symmetry. Many argued that once angular momentum was included the singularity would not form. We now know that this is not the case and, if one wishes to describe a somewhat realistic `star' that undergoes collapse, the role of angular momentum, especially for rapidly rotating bodies, while still being poorly understood, can not be neglected. 

Despite of its simplicity, the OSD model has become the foundation of black hole physics and the main reason why astrophysicists believe that black holes do form from collapse of very massive stars. Therefore, when looking at quantum gravity induced modifications to collapse scenarios, it is just natural to start from homogeneous collapse. Most models dealing with quantum corrections to black hole formation deal in fact with modifications of the above homogeneous models.

\subsection{Numerical simulations}
In order to solve Einstein's equations for collapse in more realistic scenarios that include departures from spherical symmetry and more realistic matter models one needs to resort to computer simulations. Over the years numerical relativity has grown to become one of the major players in our understanding of energetic astrophysical phenomena such as supernova explosions and binary mergers (see \cite{shapiro-book} for details). A detailed review of numerical collapse is beyond the scope of this article, therefore here we will only briefly mention some results that are important in connection with the previous discussion.
The formation of trapping horizons in collapse of polytropic fluids was first considered in 1966 by May and White
\cite{may}.
Here we would like to mention also the early works by Stark and Piran 
\cite{piran}
which considered gravitational waves emitted by the collapse of a rotating configuration in two dimensions and the work by Eardley and Smarr
\cite{smarr}
which raised the issue of the possible formation of naked singularities, a result that was later extended to axially symmetric collapse by Shapiro and Teukolsky
\cite{shapiro}.
More recently, models of three dimensional collapse of a neutron star to a Kerr black hole were obtained in
\cite{rezzolla1} using uniform rotation and in 
\cite{rezzolla2} using differential rotation for the parent star. It is also worth mentioning how the previous result has been recently extended to the case of a Kerr-Newmann black hole in
\cite{rezzolla3}.
Numerical models for gravitational collapse within GR have mostly been developed in order to produce the gravitational wave templates that are used by interferometers such as LIGO and VIRGO to detect core collapse supernovae. It is therefore obvious that all the models considered are entirely classical.
Given the importance that quantum corrected scenarios are gaining in our theoretical understanding of the final stages of collapse it is reasonable to assume that the implementation of such models in numerical simulations will play an important role in the future of black hole physics.


\section{...And Quantum bounces}\label{semiclass}

As it can easily be seen from the hypothesis of the singularity theorems, if we are to preserve the global hyperbolicity of the space-time, in order to halt collapse and avoid the formation of the singularity one has to either violate energy conditions or modify GR.
Therefore `quantum improvements' to collapse scenarios should be understood as the idea of incorporating strong gravity modifications into Einstein's equations by either modifying the geometrical side of the equations to account for effects coming from some kind of ultra-violet (UV) completion of GR or modifying the energy momentum tensor to account for the supposed behaviour of matter at high densities.
Either way, given the lack of a theory of quantum gravity, one assumes that it is possible to write these modifications within the framework of GR as additional (averaged) terms to be added to Einstein's equations. Then such terms can be brought to the right-hand side of Einstein's equations and practically treated like an effective energy momentum tensor within classical GR.
If we start from a theory of quantum gravity which reduces to GR in the weak field then we can write the field equations in the semi-classical approximation as
\be 
G_{\mu\nu}+<G_{\mu\nu}^{\rm corr}>=T_{\mu\nu} \; ,
\ee 
where the term $<G_{\mu\nu}^{\rm corr}>$ is the semi-classical correction to the Einstein's tensor coming from the ultra-violet corrections to the classical theory (remember we have incorporated the constant $8\pi k$ into the definition of $T_{\mu\nu}$).
Then by following the procedure described above we define the effective energy momentum tensor as
\be  
T_{\mu\nu}^{\rm eff}=T_{\mu\nu}-<G_{\mu\nu}^{\rm corr}> \; ,
\ee  
and proceed to study the classical general relativistic dynamics produced by such an effective matter source. A full quantum treatment of collapse models can be given for very simple cases (such as for example the one presented in \cite{haj1}) but their validity is subject to the assumptions that are made in order to fully quantize the system and may not translate to more general cases.

\subsection{A brief history of collapse models with quantum corrected interiors}

We will focus now on the interior collapsing geometry and how it can be modified in order to avoid the formation of the singularity at the end of collapse.
There are several ways that can be pursued in order to obtain the resolution of the singularity at the end of collapse via modifications of GR. For the semi-classical models considered here, each procedure corresponds to the construction of a different $<G_{\mu\nu}^{\rm corr}>$ based on a different approach towards the UV completion of gravity. Different prescriptions imply different modifications that in turn may give different collapse models. Therefore it is somehow surprising that after several years of research it seems that a unique qualitative picture is emerging from all these different approaches. We shall briefly review them here. Although not exhaustive, we hope that this survey covers most of the approaches that have been used and most of the groups that have been working on the topic.

\textbullet\ The earliest attempts were made by writing the theory for a higher order Lagrangian and considering only the leading (GR) and next to leading terms.
This is the approach that was followed by Frolov and Vilkoviski
\cite{frolov}.
In their seminal papers they studied gravitational collapse of a null shell from one loop corrections to the gravitational Lagrangian due to quantum effects, in the context of asymptotically free gravity, and found that the shell bounces at a minimum radius and re-expands towards infinity.

\textbullet\ Another early attempt to remove the singularities was investigated by Bergmann and Roman
\cite{berg}.
Noting that energy conditions may not hold for matter under the extreme conditions that develop towards the end of collapse, they used standard GR to investigate what kind of violations of the energy conditions would allow for a resolution of the singularity.

\textbullet\ A later approach developed by Hajicek considered again the collapse of a null shell within the Hamiltonian formulation of GR
(see \cite{haj1} and \cite{haj2}). 
The use of the canonical formalism allowed the author to quantize the motion of the null shell, now treated as a wave packet, to show that the quantum shell crosses the horizon in both directions. The classical ingoing shell develops into a superposition of ingoing and outgoing shell as it approaches the strong field region where classically the singularity would form. The horizon also becomes a superposition of a black hole horizon with a white hole horizon (a `grey' horizon), with the black hole prevailing at early times and the white hole prevailing at late times.

\textbullet\ Quantization of collapse of a cloud made of null fluid was considered in \cite{vaz}. The authors quantized a null dust cloud described by the Vaidya metric with an arbitrary mass distribution. At a classical level null collapse may lead to the formation of a black hole or a naked singularity depending on the specific form of the mass function. 
Unfortunately the techniques developed in \cite{haj1} cannot be applied to to a problem with many degrees of freedom, such as collapse of a spherical cloud and a different approach leads to a somehow different picture of the final stages of collapse. 
The main difference from the previous work lies in the fact that when considering a thick matter distribution instead of a shell, the singularity can be removed only in special cases (such as for example when the wave functional vanishes at the center). This poses some question of the genericity of the results obtained for collapse of thin shells. 
A similar approach was used in
\cite{vaz2}
to quantize a cloud of time-like dust particles with the surprising result that the collapsing matter may condense just outside the horizon radius to form a quasi-static object (that could be interpreted as a gravastar).
An attempt to extend the procedure to the inhomogeneous case was developed in \cite{vaz3}.

\textbullet\ The Hamiltonian formulation of gravitational collapse of a scalar field and its quantization was developed in 
\cite{hus2}.
The authors found an upper bound for the curvature as a kinematical consequence of the construction of the quantum operators. This upper bound is maintained in the dynamical scenario and thus the classical singularity is avoided. As a consequence a Planck scale remnant is left from collapse.
Similarly in
\cite{hus3}-\cite{hus3c} a program was developed to construct the fully quantum gravitational collapse of a scalar field. 
The kinematics for a spherically symmetric
quantum gravitational system was outlined in 
\cite{hus3}. 
The definition of the `quantum black hole' was developed in
\cite{hus3a},
while the Hamiltonian formulation of the system was considered in
\cite{hus3b}.
The full dynamics of the quantum gravitational spherically symmetric scalar field was considered in
\cite{hus3c}
where the quantum Hamiltonian operator was constructed.
Finally in \cite{hus3d}
the semi-classical states for the above construction were studied in a cosmological context.
Also, semi-classical states for scalar field collapse and quantum entaglement of matter and gravity in collapse were studied in
\cite{hus3e}.

\textbullet\ In \cite{kuns3} numerical simulations of gravitational collapse of a scalar field with LQG corrections showed that as the bounce occurs it is possible for the outer horizon to depart from the Schwarzschild horizon and shrink. This is related to the equations of motion for the effective source being non conservative.
Similar results were obtained in
\cite{kuns2}
where a massless scalar field was considered. In this case two horizons form. An inner horizon where the matter content `piles up' and an outer horizon which is equivalent to that of a regular black hole solution.

\textbullet\ 
The trace anomaly corrections of a scalar field theory on a given background were considered in 
\cite{arfaei1} and \cite{arfaei2}.
The corrections, treated as a backreaction, were incorporated in the collapse scenario of a thin shell in \cite{arfaei1}, while a spherical cloud of homogeneous dust was taken as the background space-time in
\cite{arfaei2}.
Similarly to previous models collapse halts before the formation of the singularity. In this case the space-time settles to a quasi-static configuration with an outer horizon (like the black hole horizon) and an inner horizon.

\textbullet\ The idea that repulsive effects may arise in the strong field regime is connected with the idea of asymptotic safety, which implies that gravity should somehow `switch off' at high densities.
There is now substantial evidence for the existence of a non gaussian fixed point for the renormalization group flow that would provide a UV completion of gravity. This is associated with an effective gravitational action with scale dependent Newton's constant 
(see for example \cite{sauer1} and  \cite{sauer2}) and implies a modification of the Schwarzschild geometry similar to that appearing in regular black hole models.
From this point of view, regular black holes in classical GR, such as the ones described in \cite{hayward}, may be seen as the UV completion equivalent of the Schwarzschild black hole and may thus originate from gravitational collapse in a semi-classical framework that  incorporates asymptotically free gravitational corrections.
Another way to understand the idea is in terms of the gravitational degrees of freedom. If there is a maximum allowed density $ \rho_{\rm cr}$ at Planck scale then as matter approaches this threshold the gravitational degrees of freedom become diluted, thus leaving a Minkowski background in the limit of $\rho\rightarrow \rho_{\rm cr}$. In this context then, the bounce is understood as a scattering process of the collapsing particles that suddenly find themselves in a gravity-free Minkowski space-time.
To model asymptotic safety one can use several approaches.
For example, as mentioned before, one can consider a running gravitational constant $G$ obtained from the renormalization group for gravity in order to describe a scale dependent effective action. Then $G$ goes to zero in the Planckian limit.
This idea was explored by Bonanno and Reuter in the context of the Schwarzschild solution 
\cite{Bonanno}
and in the context of an `evaporating' Schwarzschild solution
\cite{Bonanno2}.
The same idea was later applied to collapse in
\cite{torres1}.
Dust collapse with a varying gravitational constant $G$ depending on the density was also considered by Casadio et al.
\cite{casadio}.
Other approaches to asymptotic free gravity are based on alternative theories of gravitation, like non-local super-renormalizable higher derivative theories. The consequences for black holes and collapse of these approaches were considered for example in
\cite{modesto}
and
\cite{us2}.
A similar model was considered in
\cite{garay1}
where the bounce occurs everywhere at the same co-moving time and the expanding, white hole, solution, is described by the time reversal of the collapse model.


\textbullet\ In recent years bouncing models have gained more and more attention. The main player in the field has been Loop Quantum Cosmology. LQC applies the techniques of LQG to cosmology and big bang models. In this context it was shown that strong field corrections to the energy density inspired by a LQG treatment of the dynamics close to the big bang singularity could resolve the singularity and replace it with a bounce
(see for example \cite{lqc}).
These models are appealing because the effective energy momentum tensor takes a particularly simple form with the energy density given by the classical energy density plus a quadratic correction, that becomes important at Planck densities, and can be written as
\be \label{rhoeff}
\rho_{\rm eff}=\rho\left(1-\frac{\rho}{\rho_{\rm cr}}\right) \; .
\ee 
Such a framework was used in the description of collapse of an homogeneous scalar field in
\cite{boj3},
while in
\cite{gos2}
a similar model that classically leads to the formation of a naked singularity was considered. In \cite{gos2}, it was shown that LQG effects close to the formation of the singularity originate an outward flux of energy that dissipates away all the matter before the singularity forms. The effective energy momentum tensor is not conserved and the system exhibits a mass loss that leads to the shrinkage of the outer horizon and the geometry to become effectively Minkowski on the onset of the bounce.
Modifications of classical dust collapse were analyzed in
\cite{us1}, where it was shown that as the effective density goes to zero collapse halts and then bounces.

\textbullet\ The effective energy-momentum tensor derived from quantum corrections changes the metric in the interior space-time. It is then possible that these quantum corrections induce a change in the exterior metric as well. These modifications may mimic Hawking radiation carrying away part of the mass-energy of the bouncing interior and they may effectively be described by a quantum-corrected Vaidya space-time. Collapse scenarios with a matter outflow in the exterior were considered in
\cite{torres3}
and
\cite{torres4}
in the context of the collapse extension of the asymptotic safety inspired solutions of Bonanno et al.
\cite{Bonanno}. The qualitative picture that emerges in this case is similar to many of the models mentioned above.

\textbullet\ Finally it is worth mentioning that the resolution of singularities in gravitational collapse has been studied also in the context of alternative theories of gravity. For example singularity avoidance has been found in higher derivative, super-renormalizable theories (see for example 
\cite{us3}). Also in the context of gravity with non vanishing torsion it has been shown that spin effects may lead to avoidance of the singularity at the end of collapse
\cite{weyssenhoff}.
Within Palatini gravity the collapse of charged fluids may lead to the formation of a wormhole in place of the central singularity (see \cite{garcia2} and \cite{garcia1}).
Finally, in \cite{hus5} it was shown how modified GR affects Choptuik's mass scaling law (see \cite{chop}) observed in the final stages of collapse of a scalar field.


 From the above it can be seen that several approaches towards a treatment of quantum effects or UV completion of gravity in the strong field have been proposed and studied. It is perhaps curious to notice that most of these approaches entail a somehow similar qualitative picture of collapse: As matter reaches very high densities, repulsive effects arise which halt collapse and produce a bounce. The region with repulsive behaviour is the portion of the space-time where quantum effects can not be neglected. This region may be confined within the collapsing cloud, extend in the exterior (to reach the horizon and possibly beyond), or it could even extend to spatial infinity for a certain amount of time. 
 
 A good example is given by marginally bound homogeneous dust with LQG modifications given by the effective energy momentum tensor in equation \eqref{rhoeff}. The equation of motion \eqref{motion2} becomes
 \be 
 \dot{a}=-\sqrt{\frac{M_0}{a}\left(1-\frac{3M_0}{\rho_{\rm cr}a^3}\right)}=-\sqrt{\frac{M_{\rm eff}}{a}} \; ,
 \ee 
 which, with the initial condition $a(0)=1$, has the simple solution
 \be 
 a=\left(a_{\rm cr}^3+\left(\sqrt{1-a_{\rm cr}^3}-\frac{3}{2}\sqrt{M_0}t\right)^2\right)^\frac{1}{3} \; ,
 \ee 
where $a_{\rm cr}^3=3M_0/\rho_{\rm cr}$ is the minimum volume scale that the star reaches before bouncing. Note that the scale factor reduces to the classical solution in the limit for $a_{\rm cr}\rightarrow 0$ (which corresponds to $\rho_{\rm cr}$ going to infinity).
This example was discussed in 
\cite{us1}
without a detailed analysis of the matching and the exterior solution. Given the mass loss due to the decrease of the effective mass, it was argued that quantum effects must propagate non locally affecting the exterior horizon instantaneously (similarly to what was discussed in \cite{kuns3}). Then the quantum gravity region extends until spatial infinity and lasts for a finite interval of time centered at the time of the bounce $t_B$. Homogeneous dust is a simple non dissipative system and after the bounce the solution is readily described by the time reversal of the collapsing solution, with the black hole effectively turning into a white hole.

A few words should be spent regarding the interpretation of the parameter $\rho_{\rm cr}$. In LQG-inspired models $\rho_{\rm cr}$ can be interpreted as the Planck density. Obviously, fixing a limiting density is not the same as fixing a limiting size, like, for example the Planck length $l_{\rm pl}$. The minimum scale implied by setting a maximum value for the density may be much larger than the Planck scale. In other words, an object that can achieve a maximum density of $\rho_{\rm cr}$ will have a minimum size much larger than the Planck length $l_{\rm pl}$. Modifications of the Schwarzschild black hole via the introduction of a minimal length have been considered in the context of the quantum description of gravity at small scales
(see for example \cite{nicolini} and references therein).
However one need not limit the analysis to the Planckian regime, as there may be other factors that halt collapse before the cloud reaches the Planck density. 
In this respect, $\rho_{\rm cr}$ is just a parameter that sets the scale of the bounce. However its actual value has great importance for the phenomenological consequences of the models. Therefore, in principle, one could hope to constrain the value of $\rho_{\rm cr}$ via observations while at the same time ruling out models that do not fit with the data. In this context, for example, a bouncing scenario originating from four-fermion interaction was investigated in 
\cite{us4}
with a critical density much lower (and a critical scale much larger) than the quantum induced Planck scale. Similarly, in the context of emergent gravity, in
\cite{garay2}
it was argued that the energy scale of quantum effects that produces the bounce may differ from the energy scale at which Lorentz violations arise and the emergent gravity picture fails. In this case the range of energies in between could be described simply by quantum field theory on Minkowski space-time.
Finally, it is worth mentioning that there exist other ways to implement the avoidance of singularities. For example by requiring a limiting value for curvature invariants. Such limiting curvature proposal was explored in cosmological context for example in
\cite{markov} and \cite{markov84}.

A detailed study of the condition for the formation of trapped surfaces in the interior, following from equation \eqref{horizon}, shows that the apparent horizon also reaches a minimum value and then expands to infinity, thus reaching the boundary of the star in a finite time. What happens at this point depends on whether the quantum effects are able to propagate in the exterior. However, the horizon in the exterior exists only until the time when the outgoing matter cloud re-emerges.
The idea of horizons with finite lifespan supports the general claim that is emerging in recent years that astrophysical black holes do not posses an event horizon but `only' a trapping horizon that exists for a finite (possibly long) time
(see \cite{hawk-new}).

Most of the attempts to resolve the singularities that arise in classical collapse make use of toy models such as this one or thin shells. It is important to notice that although these are simple models allowing for the equations to be solved explicitly, it is generally believed that they retain all the crucial properties necessary to understand the fundamental aspects of the problem and the behaviour of gravity in the strong field. 
However these models are not devoid of drawbacks and understanding what they can teach us about realistic collapse of massive stars requires a careful analysis.

\subsection{The exterior geometry}

We will now briefly discuss how the modifications to the interior geometry of the collapsing cloud may affect the exterior. The dust interior of the OSD model can easily be matched to a Schwarzschild vacuum exterior. As all the matter falls into the singularity the solution tends to become globally Schwarzschild and the horizon settles to the usual event horizon. When repulsive effects are introduced the matter bounces and re-expands and the question of how the horizon in the exterior should be treated is far from trivial.
If the exterior geometry is not affected by modifications in the interior then some non trivial matching must occur at the boundary in order for the light cones to smoothly transition form an almost Minkowskian behaviour where matter is present to an inside-black-hole-horizon behaviour in the vacuum exterior. 
However, it seems possible that quantum gravity corrections are not confined to core of the collapsing cloud but can reach the weak field regions of the space-time. After the bounce the matter is outgoing and will eventually cross the horizon outwards thus destroying it. The process may be entirely symmetric in time or it may have a preferred direction, depending, among other things, on the specifics of the fluid model employed. In turn, the process may be accompanied by a transition of the black hole to a white hole solution, the time scales of which vary from model to model.

In \cite{frolov} it was assumed that the horizon remains the usual Schwarzschild black hole horizon until the expanding matter crosses it in the outgoing direction. However this interpretation is not entirely satisfactory as the light-cone structure close to the point where expanding boundary and horizon cross is not well defined.

\begin{figure}[tt]
\centering
\includegraphics[scale=0.3]{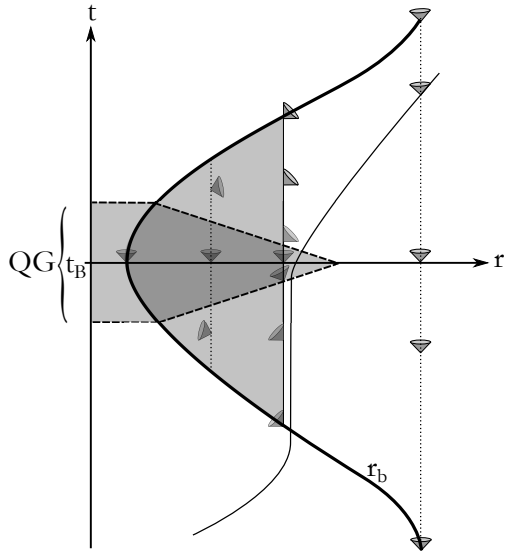}
\caption{Finkelstein diagram for the black hole to white hole transition. The grey area enclosed within dashed lines represents the region where quantum effects are important (QG). The grey area within solid lines represents the trapped region in the exterior space-time. The solid thick line $r_b$ represents the boundary of the cloud. The solid thin vertical line represents the horizon in the exterior region. 
In this case the transition is completely symmetric in time. The bounce occurs at the same time $t_B$ for all shells (as in the homogeneous case). An horizon grazing photon (thin curved line), stays in the vicinity of the horizon until right after $t_B$. The lifetime of the white hole is the same as the lifetime of the black hole.}
\label{fink1}
\end{figure}

In more recent times several researchers have proposed the idea that the exterior region may undergo a transition from black hole to white hole. In 
\cite{haj3}
the black hole to white hole transition was proposed within a model of collapsing null shells in quantum gravity without discussing the geometry of the transition. More recently, the exterior geometry induced by the transition was analyzed in
\cite{garay2} and \cite{bar3}, where it was suggested that the time scales of the transition must be short.
At the same time the effective geometry of the quantum-gravity region was studied in \cite{rov1}, where it was suggested that quantum effects may accumulate over long time scales.
In any case, models with transition from a black hole geometry to a white hole geometry can explain the change in the causal structure of the space-time by allowing for quantum effects to `tilt' the direction of the light-cones, effectively turning the black hole horizon into a white hole horizon (see figure \ref{fink1}).

As we shall see later, the lifespan of the horizon is model dependent but, for distant observers, is usually long enough to ensure compatibility with astrophysical observation of black hole candidates. At early times the models are well described by a classical GR collapse solution. Quantum corrections exist for a finite time, thus restoring classicality at late times. 


\section{Open issues}\label{open}

The general scenario presented above is extremely intriguing from an astrophysical point of view as it gets rid of space-time singularities while at the same time opening a possible observational window on quantum gravity phenomena. It is only natural then that the attention will shift from mathematical toy models to the possible features of realistic astrophysical models. 
Even though some scenarios can already be constraint by present observations, in order to truly explore the implications of these models for astrophysics one would need to have some observational data to compare against the theory.
At present, in the absence of such observational data, a way to address the physical validity of a model is to look for its internal consistency.
Most of these models present some unresolved problems or open issues that we will try to briefly discuss here.

\subsection{The horizon in the exterior}

Collapse models are constructed by tailoring a collapsing interior metric, which describes the `star', to a suitable exterior (typically Schwarzschild or Vaidya). For simplicity lets consider the case of a Schwarzschild exterior. This is the case of collapse for OSD, Lema\`itre-Tolman-Bondi (i.e. inhomogeneous dust), anisotropic fluids with only tangential pressure and fluids whose pressures vanish at the boundary. In other words, in order to have a Schwarzschild exterior there must be no inflow or outflow of matter through the boundary of the star.

In the usual OSD model when the collapsing matter passes the Schwarzschild radius a trapped surface forms at the boundary of the star. In the interior this trapped region is described by the apparent horizon, which propagates inwards to reach the center at the time of formation of the singularity. In the exterior the boundary of the trapped region corresponds to the Schwarzschild radius.
This situation may be complicated by considering inhomogeneities and more sophisticated equations of state.
For example, inhomogeneous dust collapse may result in the formation of a naked singularity, which is the consequence of a significantly different behaviour for the apparent horizon
(see \cite{joshi}).
A polytropic equation of state may also alter the structure of the horizon in the interior, delaying the time of formation of the event horizon and creating two apparent horizons (one moving inwards and one moving outwards) from a finite radius in the interior
(see \cite{musco}).
However in all these scenarios the event horizon eventually develops in the exterior and is not affected by the final fate of the collapsing matter. By looking at equation \eqref{ah} that defines the location of the apparent horizon we see that as the fluid is collapsing to the central singularity the apparent horizon also reaches the singularity. After the formation of the singularity we are left with a Schwarzschild black hole.

When repulsive effects are introduced the singularity no longer forms and the matter bounces after reaching the critical scale. This affects the apparent horizon, which also reaches a minimum value $r_{\rm min}$ and then starts moving outwards.
This can be easily seen by solving equation \eqref{ah} with the effective mass in place of the classical mass function. The existence of a minumum value for the apparent horizon indicates that an object with boundary $r_b<r_{\rm min}$ may not become trapped at all. This seems to suggest the possibility that extremely small and extremely dense compact remnants may form from gravitational collapse.
On the other hand when $r_b>r_{\rm min}$ the apparent horizon goes through an expanding phase (which may occur before the bounce) eventually reaching the boundary of the star. 
What happens there? How is the exterior affected by the bounce in the interior? Can the quantum effects reach the Schwarzschild radius and beyond?

\begin{figure}[tt]
\centering
\begin{minipage}{.45\textwidth}
\centering
\includegraphics[scale=0.3]{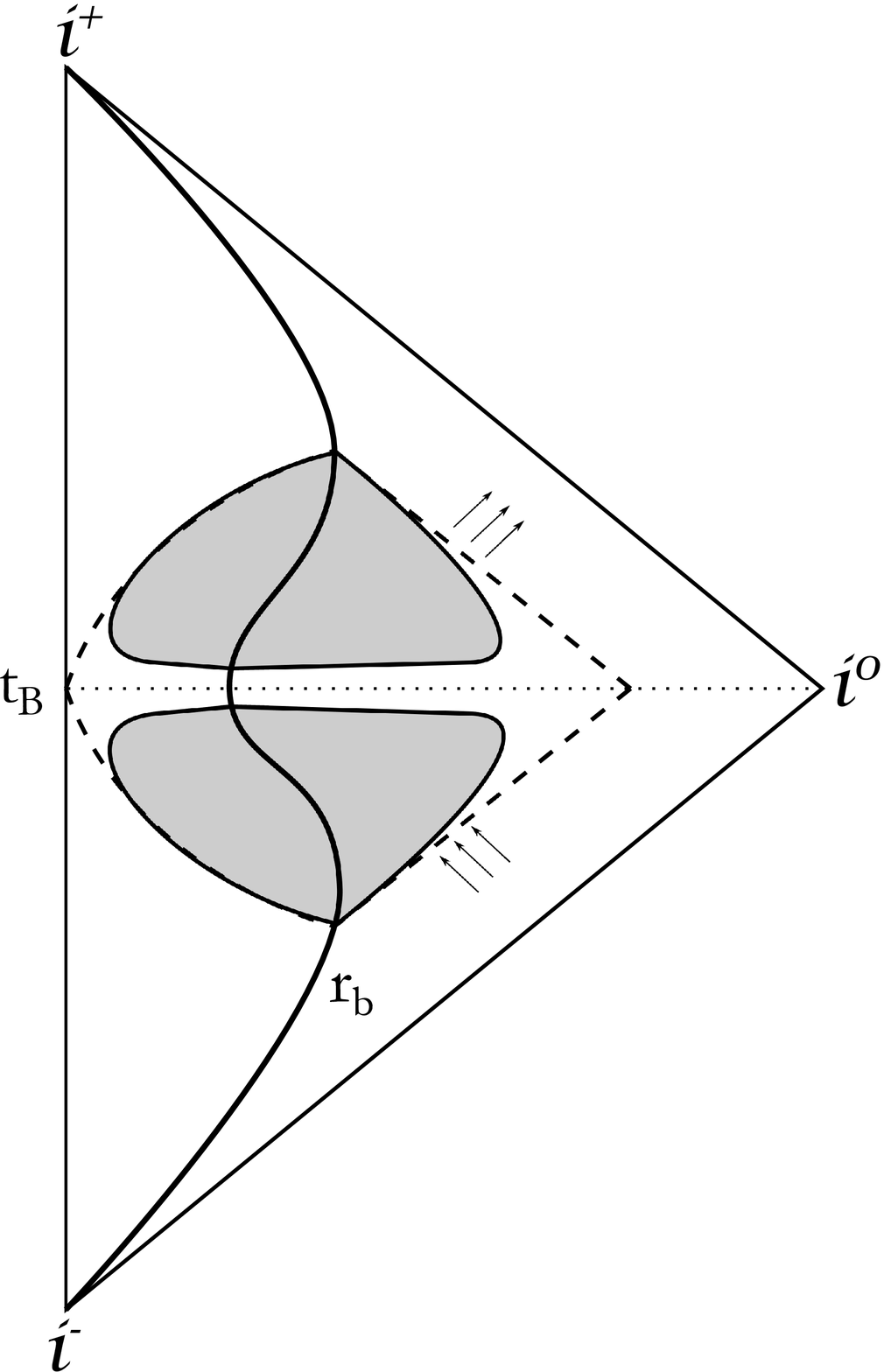}
\put(-121,93){ah}
\put(-89,93){oh}
\end{minipage}
\hfill
\begin{minipage}{.45\textwidth}
\centering
\includegraphics[scale=0.3]{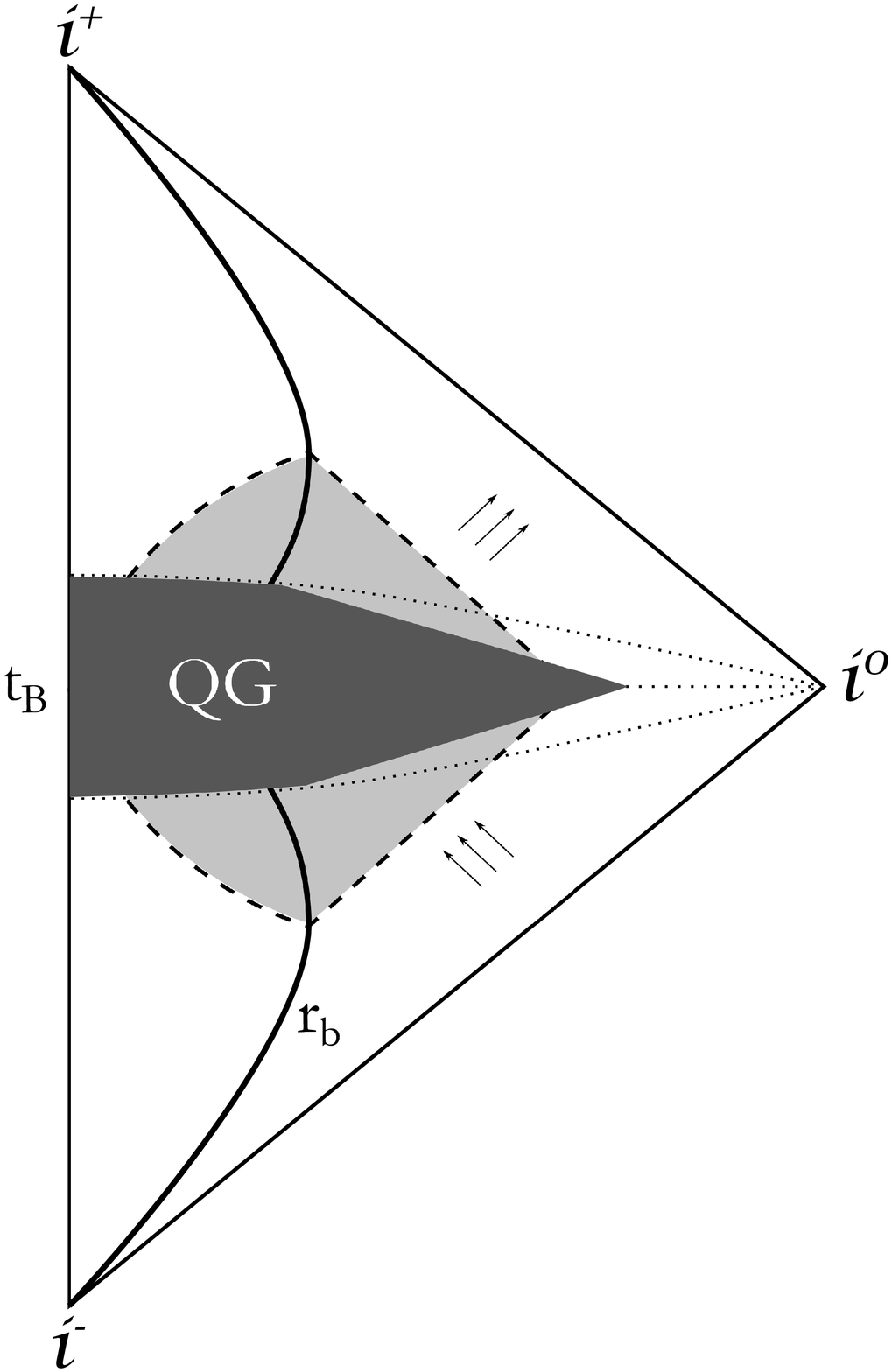}
\end{minipage}
\caption{Penrose diagrams for homogeneous collapse with semi-classical corrections. The thick solid line $r_b$ represents the boundary of the cloud. The grey areas represent the trapped regions enclosed by apparent horizons. The dashed lines represent the apparent horizon (curved) and the event horizon (straight) in the classical case. Collapse follows the classical behaviour until a certain time before the bounce. All shells bounce at the same co-moving time $t_B$. 
Left panel: When quantum effects become important the apparent horizon in the interior (ah) starts moving outwards. At the same time the outer horizon (oh) in the exterior moves inward. The two horizons meet and annihilate before $t_B$. At the time of the bounce the cloud is not trapped. After the bounce the solution is described by a time reversal of the collapsing solution, therefore a new trapped region (this time with a white hole horizon) develops for a finite time. In this scenario the bounce affects the whole space time instantaneously.
Right panel: The darker grey area represents the region where quantum effects are non negligible. Outside the quantum gravity region (QG) the space-time is given by a classical collapse solution for $t<t_B$ and its time reversal for $t>t_B$. Quantum gravity effects reach a portion of the space-time outside the horizon. Dotted lines represent lines of constant $t$ (note that due to homogeneity, in the interior, the quantum gravity region occurs at the same $t$).}
\label{bounce1}
\end{figure}

\textbullet\ One possibility is that the horizon in the exterior `feels' the repulsive effects of the inner region and shrinks to meet the outgoing apparent horizon at the boundary
(see left panel in figure \ref{bounce1}).
This scenario is easily illustrated in the simple example of homogeneous dust. The effective mass becomes zero at the time of the bounce, thus making the space-time almost Minkowki. In fact, at the time of the bounce $\rho_{\rm eff}=0$ and $p_{\rm eff}=-\rho_{\rm eff}$ and from equation \eqref{ah} with $M_{\rm eff}$ in place of $M$ we see that there are no trapped surfaces in the interior at $t=t_B$. One can see that there may not be trapped surfaces in the exterior as well if this is described by a Vaidya radiating solution with $M(v)$ obtained from $M_{\rm eff}$. Then the horizon in the exterior is affected instantaneously by the bounce in the interior. This is reminiscent of the Newtonian problem of action at distance and could be understood if the horizon was in some way `entangled' with the infalling matter. In this case, the changes in the effective energy-momentum tensor are felt immediately everywhere in the space-time and so the horizon in the exterior shrinks due to the decrease of the active gravitational mass in the interior. The two horizons eventually `meet' again at the boundary at a time before the bounce. In this case the physics of the bounce is not covered by an horizon and the quantum gravity region could in principle have an observable signature visible to far away observers (see for example
\cite{us1}).

\textbullet\ Another possibility is that the exterior region is not described by the classical Schwarzschild solution but by a modified (`quantum corrected') solution in the form of a regular black hole space-time. Regular black holes can be constructed under several prescriptions and share some qualitative properties. Given the absence of the singularity, regular black holes may present multiple horizons (when the mass exceeds a critical mass), the simplest situation being that of two horizons, an outer horizon close to the classical Schwarzschild horizon and one inner horizon. The outer horizon is in the weak field region and quantum effects have little impact on its structure. On the other hand the inner horizon is a distinctive feature of the `quantum' corrections. 
For example in
\cite{hayward}
it was considered a very simple metric for a regular black hole inspired by \cite{bardeen}
and it was shown that the horizons in the modified space-time is located at the zeros of
\be 
f(r)=1-\frac{2Mr^2}{r^3+2l^2M} \; ,
\ee 
where $l$ is a characteristic length that can be thought to be of the order of the Planck length. Then for $M>(3\sqrt{3}/4)l$ the solution has two horizons, as discussed above. Note that for $l=0$ one retrieves the Schwarzschild solution.
In the dynamical case, if the exterior is described by a regular black hole solution with two horizons, the outgoing apparent horizon within the star may match with the inner horizon of the exterior metric once it crosses the boundary, while the outer regions of the black hole remain mostly unaffected until the time of the bounce.

\textbullet\ The bounce of the apparent horizon is also connected to the role of the effective mass $M_{\rm eff}$.
Consider again for simplicity the homogeneous dust case. In evaluating the physical density we must use the physical mass function $M_0$ and thus we obtain $\rho=3M_0/a^3$, which, with the `quantum corrected' scale factor, reaches the maximum value $\rho_{\rm cr}$ at the time of the bounce. If we use the physical mass function instead of the effective mass function in the equation for the apparent horizon \eqref{ah} we obtain the intriguing situation where the apparent horizon reaches the smallest radius at the time of the bounce and then re-expands.
This allows for the exterior to have only one horizon that behaves like the event horizon in the Schwarzschild space-time (plus possible small corrections) until the bounce (or until it is reached by the effects of the strong gravity region). After the bounce the model is again described by the time reversal of the collapsing scenario.
Unfortunately this situation doesn't seem to be justifiable if we consider the geometrical definition of the apparent horizon. In fact, the condition for the formation of trapped surfaces in the interior is given by
\be 
g^{\mu\nu}(\partial_\mu R)(\partial_\nu R)=0  \; ,
\ee 
which is equivalent to requiring that the surface $R(r,t)={\rm const}.$
is null. In the case of the semi-classical model for collapse this implies equation \eqref{ah} with $M$ replaced by the effective mass $M_{\rm eff}$.

\textbullet\ In any case it would seem that the exterior geometry cannot remain unaffected by the bounce and thus the quantum effects must somehow reach the horizon in the exterior and possibly beyond. 
We can suppose that reaching a certain threshold density may signal when the cloud is entering the quantum gravity regime. In homogeneous collapse every shell has the same density and therefore quantum gravity effects will start throughout the cloud at the same time.
After that time the effects may reach the boundary and propagate in the exterior to modify the geometry near the horizon (see right panel in figure \ref{bounce1}).
In the model discussed in
\cite{us1}
quantum effects reach arbitrarily large distances at the time of the bounce.
On the other hand it is possible to construct models, such as the one discussed in
\cite{rov1},
where quantum effects reach a finite distance beyond the horizon and the space-time is always classical at large distances.

The most striking consequence is that quantum effects may alter the geometry of the space-time in a region where classical physics should dominate. This shows better than any examples how the black hole horizons are intrinsically quantum gravitational concepts.
Another thing that stands out from the previous considerations is that the evolution after the bounce can be described as the time reversal of the collapse scenario. If this is true both in the interior and the exterior then the black hole turns into a white hole as the horizon turns from a membrane that doesn't let anything escape to a membrane that doesn't let anything enter. From the above considerations one is led to ask what is the mechanism that leads the physics of the strong gravity region to have effects at large distances, or how the quantum effects propagate to reach the horizon. And how such effects can turn the black hole into a white hole.

\subsection{The black hole to white hole transition}\label{trans}

In this section we shall investigate the question of how quantum effects may reach the horizon. The simplest mathematical model where the evolution after the bounce is described by the time reversal of the collapse model does not address how the black hole event horizon turns into a white hole horizon. 
The event horizon in the exterior region is within the regime where gravity is well described by GR and relatively far from the central region where matter is confined just before the bounce. Therefore if the bouncing scenario induces a transition of the black hole to a white hole solution it is important to understand how such a process can happen. How effects that occur in the strong gravity region affect the geometry near the horizon.

Three main possibilities can be suggested: The horizon could change nature as a consequence of some signal propagating from the strong gravity region at a finite speed. Or the transition could be a statistical process with the black hole horizon being the limit at early times and the white hole horizon the limit at late times, with a `grey' hole given by a superposition of the two in between. Finally it could be that the horizon is quantum entangled with the ingoing matter and therefore reacts instantaneously to what happens close to the center.


\textbullet\ If some signal propagates from the quantum gravity region outwards to reach the horizon in a finite time then the horizon in the exterior is a black hole horizon until it is affected by the signal. Then it is worth wondering about the nature of the signal and what kind of carrier would propagate it towards the exterior.
For example, in 
\cite{bar2}
one such mechanism was suggested in the form of a shock wave originating near the strong curvature region.
In the time symmetric case, the quantum gravity region reaches the horizon just before the time of the bounce and the black hole turns quickly (for co-moving observers) into a white hole. For example in the models described in \cite{haj1} the superposition state lasts for a very short time. The geometry before the transition is well described by the classical black hole space-time, while after the transition it is a classical white hole solution (the case of a fast transition from black hole to white hole is illustrated in figure \ref{fink1}).

\textbullet\ On the other hand it is possible that quantum effects accumulate over time in the outer regions making the transition a statistical phenomenon. 
This is equivalent to say that at large scales (near the horizon) quantum effects are negligible at any given time but they pile up eventually leading the solution away from the classical black hole solution
\cite{rov1}.
In this case there are a few questions that are worth asking. What is the nature of the horizon during the transition phase? How long does the transition last before the horizon settles to a white hole horizon? And when does the transition occur? Close to the time of the bounce (thus making the scenario symmetrical in time) or close to the disappearance of the horizon (thus making the white hole phase short lived?
In this case it is possible that the transition during which the horizon is a superposition of black hole and white hole lasts for a long time even for local observers near the horizon thus having different observable effects with respect to the previous scenario. Quantum effects would start to accumulate at the horizon right after its formation and the transition would affect the whole lifetime of the black hole.
Note that the two situations described above offer the possibility to break the time reversal symmetry if the quantum gravity signal reaches the horizon after the bounce, as we shall see later. 

\begin{figure}[h]
\centering
\begin{minipage}{.45\textwidth}
\centering
\includegraphics[scale=0.3]{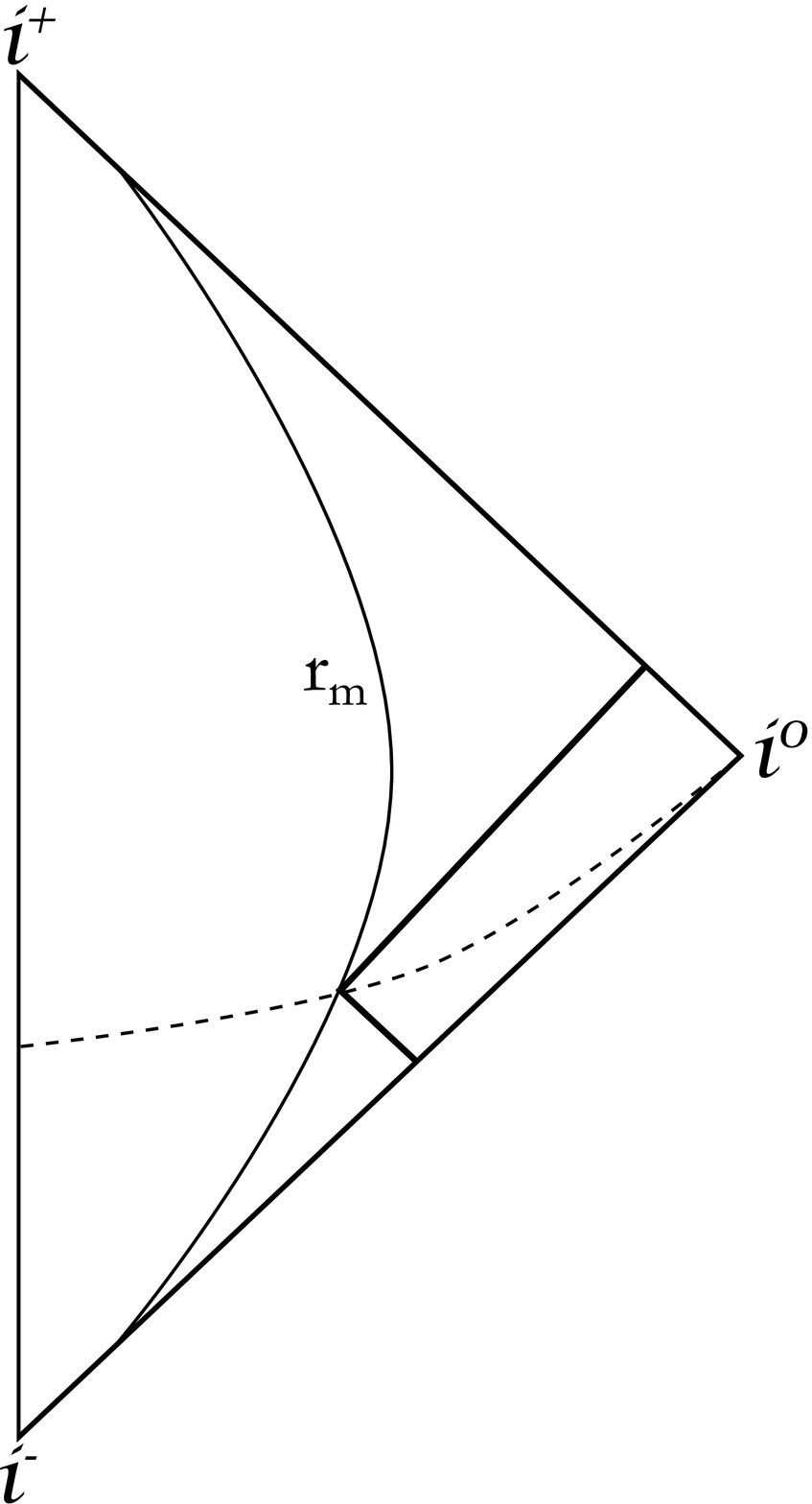}
\end{minipage}
\hfill
\begin{minipage}{.45\textwidth}
\centering
\includegraphics[scale=0.3]{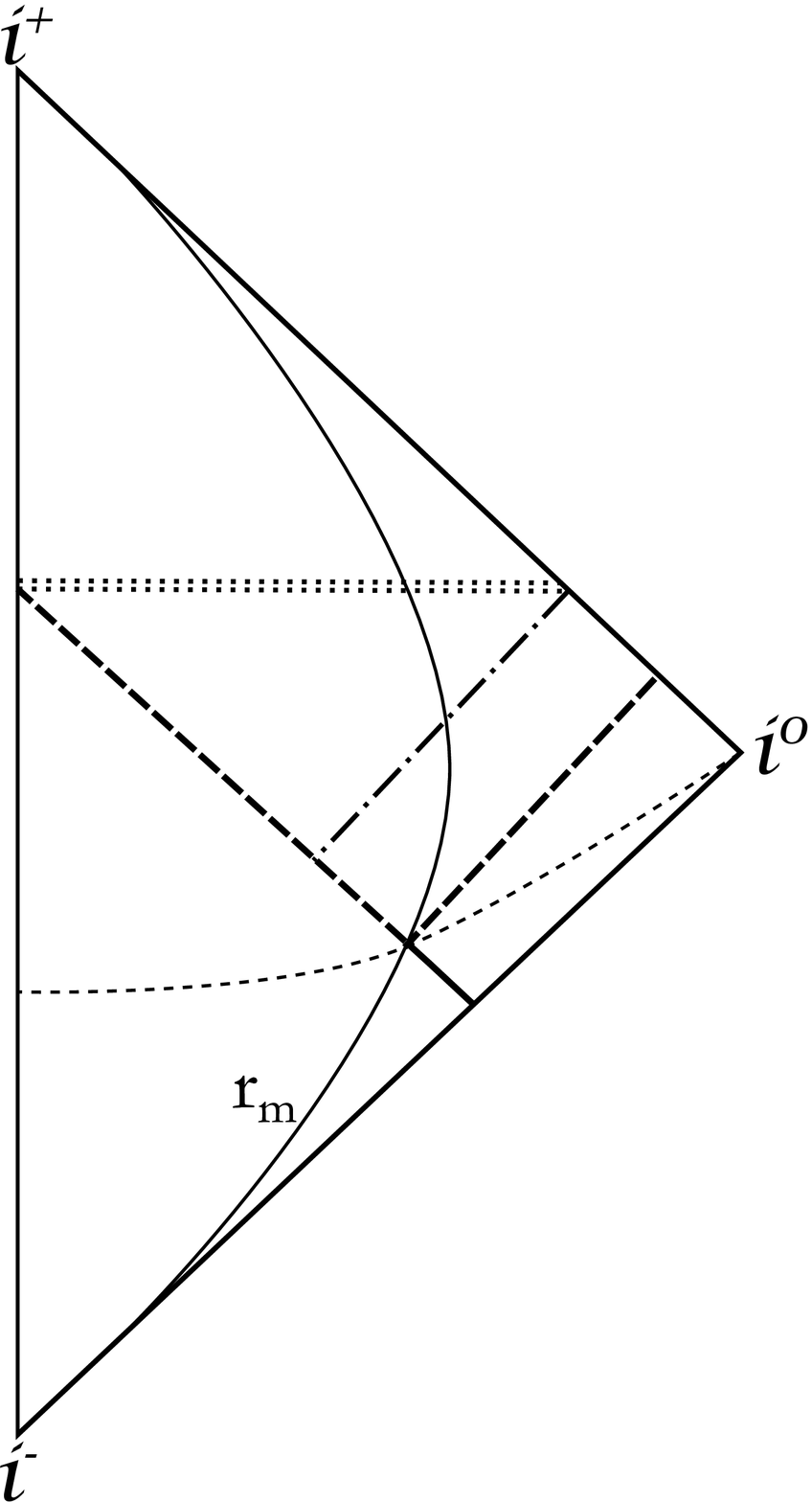}
\end{minipage}
\caption{Penrose diagrams of a collapsing Vaidya null shell with a massless mirror (left panel) and a semi-reflective mirror (right panel). The trajectory of the static mirror is given by the fixed value of the radius $r=r_{\rm m}>2M$ (solid thin line). The dotted line represents a line of constant $t$.
Left panel: The ingoing wavepacket (thick solid line) reaches the mirror and bounces back to infinity.
Right panel: The ingoing wavepacket reaches the semi-reflective mirror. The wave function is split into two entangled parts, one ingoing and one outgoing (dashed lines). As a consequence, when the ingoing part crosses the Schwarzschild radius, an `entangled' horizon (dot-dashed line) and eventually an `entangled' singularity (double dotted line) form. 
If a detector is placed on the trajectory of the outgoing part of the wavepacket the whole system will collapse to either the diagram in the left panel of this figure (if there is detection) or to the diagram in the right panel of figure \ref{fig1} (if there is no detection).}
\label{vaidya}
\end{figure}

\textbullet\ It is also possible that the horizon is in some way `entangled' with the infalling matter that produced it and therefore responds instantaneously to quantum effects from the vicinity of the center.
This third possibility corresponds to asking the question `Is it possible that faster than light communication occurs between the quantum gravity region in the dense core and the horizon?'
In order to address this point let's consider a simple thought experiment based again on collapse of a Vaidya null shell. Following \cite{haj1} we assume that the null dust shell can be quantized and described by a wave function. Let's imagine to position a massless spherical mirror at some fixed radius $r_{\rm m}>2M$. Without the mirror and without a quantum treatment, the shell would collapse and form a singularity (as shown in the right panel of figure \ref{fig1}). With the mirror the shell would bounce before crossing the horizon radius and re-expand to infinity (see left panel of figure \ref{vaidya}). With the quantum treatment and without the mirror, the shell would bounce close to the center following the dynamics described in \cite{haj1} and create a state of superposition between black hole and white hole. Now imagine to position a semi-reflective mirror at a fixed radius $r_{\rm m}>2M$. With the semi-reflective mirror, we can in principle follow the usual framework of quantum interference and conclude that the wave function should split into two parts, one transmitted, that propagates towards the center and one reflected, that propagates towards radial infinity. An observer at a radius $r>r_{\rm m}$ would have a probability of $1/2$ to detect the expanding shell. Then even before the white hole transition due to the bounce occurs, the geometry would be described by a superposition of two geometries (one without the horizon and one with the usual black hole horizon), showing that quantization of the null shell entails quantization of the horizon even in the weak field (see right panel in figure \ref{vaidya}).
In this kind of scenario the transition from black hole to white hole would be instantaneous, as the horizon, being entangled with the collapsing matter, would change its nature at the time of the bounce.

In all of the above cases it seems unavoidable that the transition from black hole to white hole must have a non causal nature. In fact the light cones in the exterior region but inside the Schwarzschild radius are tilted towards the center while the quantum gravity effects are propagating towards outer radii. This can be understood in terms of the effective energy-momentum tensor, as violations of the dominant energy condition allow for signals to propagate outside the light cone.

It is worth noting that in \cite{bar2} it was argued that the time-scales of the transition must be short for the white hole to be stable against perturbations.
The model for collapse of quantum null shell developed in 
\cite{haj3} 
supports the idea that the transition must be short. Later work by Barcel\'o et al.
also supports this claim
(see \cite{bar3}).
At the same time, if the transition is a statistical effect, then the timescales might become relatively long.
This naturally leads to the question of the lifetime of the horizon in these bouncing scenarios. The time scale of duration of the transition may be short while the observed lifetime of the black hole for distant observers may still be long (million of years for a stellar mass object). Can distant observers see the state of superposition of black hole and white hole? Or do they see a sudden transition? And how long does the black hole horizon live for distant observers? 

\subsection{Lifespan of the black hole}

The Schwarzschild black hole exists forever. In the OS model after all matter has reached the singularity the final product of collapse is a Schwarzschild black hole. The event horizon in the Schwarzschild geometry `knows' the future development of the space-time for `the entire history of the universe'. It has been argued several times that event horizons can not be detected and are not relevant for real physical phenomena happening in the universe
(see for example \cite{vis-hor} and references therein).
The event horizon is not a physically observable feature of the space-time. On the other hand, from the point of view of local experiments, apparent horizons, which have local significance, can be observed. Therefore when talking about black holes in these bouncing models one has to think about a trapped region, enclosed within apparent horizons, that exists for a finite, albeit possibly very long, time.

In bouncing models, the black hole horizon has a finite lifespan that ends when either the black hole turns into a white hole or when the expanding matter crosses the horizon outwards. Then for how long does the black hole `live'? Is the lifespan of the black hole horizon as measured by distant observers compatible with current observations of astrophysical black hole candidates?
If the lifetime of these objects is short enough then there is a possibility that the observational signature of the disappearance of the horizon be experimentally tested by modern telescopes.
If, on the other hand, the time scales for the disappearance of the horizon are long (say comparable with the age of the universe), then there is little hope to impose observational constraints on the theoretical models.
In some cases the lifespan of the black hole can be calculated using the usual arguments found for example in
\cite{MTW}.
On the other hand the same calculation in a fully quantum framework becomes much more complicated.

\textbullet\ For example, in \cite{frolov} it was found that the lifetime of the black hole is of the order of $T\simeq Me^{M/m_{\rm pl}}$, where $M$ is the mass of the collapsing shell and $m_{\rm pl}$ is the Planck mass (obviously for $m_{\rm pl}\rightarrow 0$ the lifetime goes to infinity). The fact that the lifespan is exponential in the Schwarzschild mass makes it longer than the age of the universe for a stellar mass black hole.

\textbullet\ However, the tunnelling process suggested in \cite{haj3} and the transition process suggested in \cite{bar2} indicate a time scale of the order of $T\simeq t_{\rm pl}(M/m_{\rm pl})$, where $t_{\rm pl}$ is a model dependent characteristic time scale that can be thought of as related to the Planck time. Arguments for a short lived horizon were suggested for example in
\cite{bar3}
where a short transition from black hole to white hole implied also a short lifespan for the black hole horizon (of the order of milliseconds for distant observers).
Using a different framework, in 
\cite{bar4} 
the authors evaluated the mean lifetime of the black hole by computing the probability of the same of tunneling to a white hole solution. They found an exponential decay law of black holes to white holes with a mean lifetime that scales like $1/(4M)$, which is small for stellar mass objects. When this factor is added to the time that is required for the star to bounce as seen by distant observers the striking result is that quantum tunneling has little effect. The decay does not modify significantly the lifetime of the black hole for far away observers, which scales like the Schwarzschild mass $M$.

\textbullet\ On the other hand, the tunneling model described in 
\cite{rov1}
has a time scale of the order of $T\simeq M^2/l_{\rm pl}$, where again $l_{\rm pl}$ is a model dependent characteristic length that can be thought of as related to the Planck length (note that in geometrized units we would have $m_{\rm pl}=t_{\rm pl}=l_{\rm pl}=\sqrt{\hbar}$). The fact that the lifespan of the black hole is quadratic in the Schwarzschild mass puts this model somehow in between the two models presented above, shorter than Hawking evaporation scales (which are of the order of $M^3$) but longer than the time scales discussed above.
The same estimate was obtained within LQG in 
\cite{rov-new},
where the authors suggested the possible existence of new astrophysical phenomena, given the solid theoretical support for black holes tunneling to white holes in sufficiently short time scales. The idea put forward is that Fast Radio Bursts may be the result of moon-sized primordial black holes exploding today.

In general there are three phases that the outer horizon goes through before the outgoing matter re-emerges (black hole phase, transition phase and white hole phase). The black hole horizon lasts until the transition to white hole begins. The transition may be instantaneous, short or long and produces what Hajicek called a `grey horizon', which can be thought of as a superposition of the other two states. After the transition the system settles to a white hole horizon.
The lifespan of the black hole, strictly speaking, is the time, as measured by distant observers, during which the black hole horizon exists.
It should be noted that in time symmetric models a long lived black hole horizon implies a long lived white hole horizon which may be problematic due to the known instabilities of white hole solutions
(see for example \cite{white1} and \cite{white2}). 
One easy way to understand this instability is to think of the white hole as an attractive body with repulsive effects. Matter is still attracted towards the center due to the positive mass trapped within the white hole horizon, however it can not pass the horizon due to the nature of the white hole horizon itself, therefore it is forced to accumulate at a finite distance from the horizon thus eventually forcing a new collapse.
As we shall see, dissipative effects and inhomogeneities may help solving this problem.

\subsection{Hawking radiation and time symmetry}

It is interesting to compare the previous results with the Hawking evaporation time, which is of the order of 
\be 
T\simeq t_{\rm pl}\left(\frac{M}{m_{\rm pl}}\right)^3  \; ,
\ee
thus considerably longer that any quantum-corrected bouncing scenario. This suggests the possibility that the quantum processes that govern the black hole to white hole transition may super-seed the effects of Hawking radiation on astrophysical black holes.

\textbullet\ Hawking radiation: When considering semi-classical effects in the vicinity of the event horizon one finds that particle pairs created from vacuum fluctuations may cause the black hole to gradually lose its mass through what is now called the Hawking evaporation process \cite{hawking-radiation}. Hawking radiation is believed to be a real physical process that happens in the vicinity of the black hole horizon and must be taken into consideration if one aims at understanding the physics of black holes. Also it has been noted by several authors that the regularization of the Schwarzschild singularity due to UV effects in the strong field has implications for the evaporation process and the information loss problem
(see for example \cite{paradigm}).
Therefore, in order to improve on the physical validity of collapse model several authors have included Hawking radiation from the horizon after collapse passes the threshold of the Schwarzschild radius
(see for example 
\cite{rad1} and \cite{rad2}). 
Furthermore, it has been shown that back reaction from Hawking radiation in gravitational collapse may in fact halt collapse without the need for other repulsive effects
\cite{mersini}.
Therefore it would seem that Hawking radiation is an important effect that must be considered in any discussion of the dynamical processes that lead to black hole formation.
 For example in
\cite{torres2}
a generalized Vaidya space-time was used to model Hawking radiation in the exterior of the collapsing sphere. This approach has some useful advantages as it allows to include radiation in the geometry of the exterior space-time.
However, modeling Hawking radiation with an outgoing Vaidya metric might not accurately capture the main features of the process.
The outgoing Vaidya solution describes what can be regarded as a null fluid which is somehow different from the particle flux of Hawking radiation.
Also the outgoing radiation would be coming from the boundary surface of the collapsing object and not from the vicinity of the horizon as required by semi-classical vacuum fluctuations. 

\textbullet\ The black hole `atmosphere': Most models that consider Hawking evaporation treat the radiation as coming from a region very close to the horizon and having a measurable temperature. These two points, together with the time scales for evaporation, are in fact very important when considering Hawking radiation from astrophysical black holes.
Does Hawking radiation come from the vicinity of the horizon?
In 
\cite{gid2}
it was shown, using Stefan-Boltzmann law, that despite being a semi-classical effect, Hawking radiation appears quite far from the horizon, from an effective radius of the order of $3\sqrt{3}M$, considerably larger than $2M$. Therefore one has to understand Hawking radiation as coming from a `quantum atmosphere' that extends well beyond the horizon. 
The flux of particles coming from this atmosphere peaks around $4M$
\cite{liberati},
thus confirming that considering Hawking radiation as a near horizon effect is a mistake.
If one considers the thermal wavelength $\lambda_T$ of the emitted radiation one gets a value of the order $\lambda_T\approx 158M$, much larger than the horizon. The peak wavelength for Hawking radiation from a stellar mass black hole would then be of the order of $10^5 {\rm m}$, much longer than the cosmic microwave background (CMB). This also suggests that astrophysical black holes, being immersed in a bath of photons from the CMB, may not be radiating at all.

So, is Hawking radiation a necessary ingredient for a realistic collapse model? These considerations, together with the fact that Hawking evaporation takes time scales longer than the age of the universe in order to completely evaporate a stellar mass black hole, suggest that the evaporation process might be entirely overpowered by the black hole to white hole transition mechanism.

\begin{figure}[tt]
\centering
\includegraphics[scale=0.3]{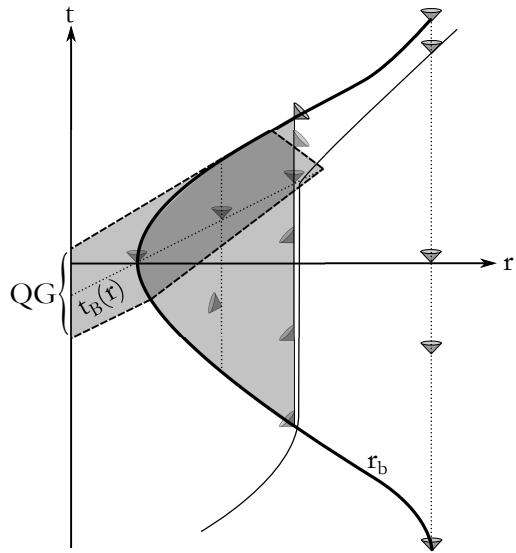}
\caption{Finkelstein diagrams for the black hole to white hole transition with asymmetry in time. The grey area enclosed within dashed lines represents the region where quantum effects are important (QG). The grey area within solid lines represents the trapped region in the exterior space-time. The solid thick line $r_b$ represents the boundary of the cloud. The solid thin vertical line represents the horizon in the exterior region. 
In the time asymmetric transition, the bounce curve $t_B(r)$ (dotted line) is not constant (see figure \ref{fink1} for comparison with the symmetric case). An horizon grazing photon (thin curved line) remains in the vicinity of the horizon for longer time with respect to the symmetric case. The lifetime of the white hole is short as compared with the lifetime of the black hole.}
\label{fink2}
\end{figure}

\textbullet\ Time symmetry:
It should be noted that Hawking radiation does not turn the black hole into white hole and so the process may `afford' to have such a long timescale as it does not clash with white hole instabilities.  The Hawking evaporation process, being an intrinsically dissipative process, possesses a clear direction in time and its implications for the connection between thermodynamics and black holes are well known
\cite{thermodynamics}.
On the other hand, the toy models of black hole to white hole transition are entirely time symmetric, as they neglect any dissipative effects, leaving open the question whether more realistic evolution models may show some asymmetry in time and how this asymmetry would manifest. 
There are several indications that the black hole to white hole transition may not be time symmetric. The symmetry may be a consequence of the initial assumptions (in some cases the bounce is obtained through a `cut and paste' technique of the metric with its time reversal across the curve describing the time of the bounce).
For example, consider the hypothetical scenario in which the strong-field effects reach the horizon after the bounce. In this case the black hole horizon may be long lived while the white hole horizon may last only for a short time. A situation like this may be triggered by the presence of inhomogeneities in the collapsing cloud.
In this kind of scenario the transition should again occur over a short time thus possibly leaving the white hole solution as a short lived unstable state that quickly `explodes' releasing the matter that was previously trapped (see figure \ref{fink2}).
The asymmetry has the added benefit to cure the problem of white hole instability, since the white hole state does not last for long enough to become unstable. For example in 
\cite{perez}
a similar trick was employed to resolve the white hole instability of the fireworks model presented in
\cite{rov1}.

Another, obvious, indication that the bounce should not be symmetric in time comes from the second law of thermodynamics. Hawking radiation is a dissipative effect, but it is not the only one. In a realistic fluid, entropy is expected to increase in time regardless of whether matter is collapsing or expanding.  On the other hand in the time symmetric models described above after the bounce the matter cloud comes back to the exact initial configuration, thus with its entropy unchanged (in fact most models consider fluids with constant entropy). In these models if the entropy increased during collapse it must decrease after the bounce thus contradicting the second law of thermodynamics. This is just a consequence of the too simplistic matter profiles that have been chosen. 
Therefore, the issue of the thermodynamics of the bouncing models is closely related to the specific matter models used to describe them. Obviously, properties of matter in the strong field, such as equation of state and energy conditions, play an important role in determining the final fate of collapse. 

\subsection{Matter models}


In this section we will discuss the features that matter models are expected to exhibit in the strong field. 
Non dissipative, homogeneous and isotropic matter fields are not realistic. How does the bounce scenario changes once we introduce, for example, inhomogeneities?
Also, the usual weak field equations of state are not expected to hold in the strong field. What can we say about the properties of matter in the strong field?

\textbullet\ Inhomogeneities: We mentioned that in homogeneous dust collapse every shell has the same density and therefore quantum gravity effects will appear throughout the cloud at the same co-moving time.
It is reasonable to assume that inhomogeneities will affect the collapse scenario just as much as they do in the classical case. For example, in classical collapse introducing inhomogeneities in the dust model changes the location of initial formation of the horizon from the boundary to the interior of the cloud (see \cite{booth} or \cite{may} for earlier numerical work on polytropic fluids). What happens when semi-classical corrections are considered?
Since the critical density now is reached by different shells at different time, it seems natural to suppose that also the bounce will occur at a different time for each shell. The time of the bounce can then be described by a curve $t_B(r)$
(see figure \ref{fink2}).
Then some shells will be collapsing as some others expand, thus breaking the time symmetry of the homogeneous model.
Note that in collapse scenarios, as opposed to cosmological models, it is possible to introduce inhomogeneities in a natural way since we begin with a classical configuration where the behaviour of the matter fields is well understood. On the contrary, in cosmological models one has to introduce the fluctuations initially at a quantum level. It is worth noting that some scenarios considering the effective dynamics of inhomogeneous cosmological models within LQC have been studied (see for example
\cite{Mena1} and \cite{Mena2}).

Considering a decreasing density profile the critical density is reached first at the center of the cloud, thus causing the inner shells to bounce first and collide with the ingoing outer ones 
\cite{yue}.
In this case it is natural to ask what happens when expanding shells and collapsing shells collide. It is possible that shell crossing singularities develop thus causing caustics and shockwaves that may disrupt the entire collapse. This process is not time reversable. Considerations along these lines were discussed in 
\cite{yue}
where the authors introduced inhomogeneities at a perturbative level, close to the center of the collapsing sphere. In
\cite{harada}
the authors used a non perturbative approach to the introduction of inhomogeneities and found numerical indications that the collapse halts in a shell-crossing singularity before the formation of the central singularity.
It is clear that the exact form of the density profile plays a crucial role in the future evolution of the cloud. Also, if shell crossings do arise at some stage, then the evolution after the bounce will not be the time reversal of the pre-bounce scenario.

All the models discussed here are extremely simple (dust, homogeneous perfect fluids and null dust shells). More realistic models such as polytropic fluids could be considered in the weak field and this would naturally lead to complicated sets of equations that need numerical tools to be solved. 

\textbullet\ Energy conditions: 
Every known classical matter field describing macroscopic objects obeys energy conditions, such as the `dominant energy condition' which roughly states that energy momentum cannot travel faster than light. The least stringent of the energy conditions however is the `weak energy condition'. The weak energy condition essentially states that the energy density, as measured by an observer moving on a time-like curve, must be non-negative. 
However, when dealing with matter fields over very short distances, one has to take into account effects coming from quantum field theory. Then it is possible that the expectation value of the energy density becomes negative. Furthermore, it is possible to make such expectation value arbitrarily negative thus having a matter field that violates the weak energy condition.
Therefore it is clear that classical energy conditions must be modified in situations where one has to include quantum effects. After all, energy conditions are constructed to serve a purpose. That is to describe the behaviour of realistic matter fields under some assumptions. Therefore if realistic matter fields are found to violate energy conditions, no harm comes from dropping them or replacing them with more suitable versions
(see for example the case of the `trace energy condition' discussed in
\cite{visser}).
Then it is legitimate to ask how we should modify such energy conditions to account for ultra-violet effects in the strong gravity regime. 

Quantum inequalities (see
\cite{ford})
were introduced to constrain the negative densities in quantum matter fields. Energy conditions are conditions imposed on the energy momentum tensor in order to constrain its physical validity. In the same way, quantum inequalities, imposing constraints on the semi-classical energy-momentum tensor, allow for checking the physical validity of effective matter models.
Similarly, the flux energy conditions
\cite{FEC}
and its quantum equivalent
\cite{FEC2}
were introduced to impose local semi-classical constraints on the energy momentum tensor.
In \cite{visser2}
some properties and applications of the flux energy condition and its modifications were studied, and their possible ranges of applications were discussed.

At present, what kind of energy conditions will hold for matter fields in the strong gravity regime is still unknown. This uncertainty is deeply related to our present lack of understanding of the behaviour of matter at ultra-high densities.
As said, it is reasonable to assume that the usual polytropic and barotropic equations of state will not hold at the densities found at the core of neutron stars or in collapse close to the formation of black holes. But then what kind of equation of state can suitably describe matter fields in these regimes?

\textbullet\ Equations of state: Equations of state for high density matter have been a topic of study for decades. It is indeed possible that some kind of exotic matter is already present at the core of neutron stars. Quark matter seems to be an ideal candidate and it is indeed possible that new islands of stability may appear below the neutron star degeneracy pressure once matter reaches such a state
\cite{weber}.
Historically there have been two kinds of equations of state considered to describe matter at high densities, and they are on the opposite sides of the spectrum when it comes to their physical properties. 
On one hand, there are hard equations of state, like the stiff fluid model with $p\rightarrow \rho$ as densities increase. This possibility was advocated for example by Zel'dovich
\cite{zeld}
and entails, among other things, a sound speed within the cloud that approaches the speed of light. 
On the other hand, it has been argued that asymptotic safety and the decreased effects of gravitation at high densities might imply the opposite behaviour for matter. Namely matter fields could tend towards a soft equation of state, where the speed of sound in the cloud decreases as densities increase.
This idea was already suggested by Sakharov in 
\cite{sakh}
where it was argued that as the baryon density increases the energy density may in fact become lower.
A similar idea was advocated by Hagedorn, who suggested that as densities increase the number of allowed states saturates
and the system tends to a limit phase with a maximum temperature 
\cite{hagedorn}.
As matter approaches the Hagedorn limit, adding energy
to the system does not increase the temperature but instead creates more and more particle pairs. One can think of a system near the Hagedorn phase as allowed to store any arbitrary amount of energy without increasing its temperature. Consequences of such a state of matter for cosmology and collapse were explored in
\cite{bahcall}.
More recently, semi-classical models for collapse of Hagedorn fluids were studied in
\cite{me}
and
\cite{harko}.
In this context for example, it was shown that if there is a limiting density threshold in nature that can not be passed by any kind of matter then it is possible to rewrite the equations of GR in a form that reduces to a DeSitter (read, cosmological constant) space-time in the limit
\cite{markov}.

\subsection{Other possibilities}

\textbullet\ Baby Universes: Another intriguing possibility is obtained if one assumes that the exterior horizon is well described by the classical event horizon and that the bounce of the matter is confined within the black hole. In such case far away observers would see collapse evolving as in the standard relativistic picture, with the formation of the horizon and the system settling down to a Schwarzschild black hole. However the infalling matter, instead of being condemned to falling into the singularity, would bounce and re-expand within the confined environment of the newly formed black hole thus originating a baby universe. 

\begin{figure}[ht]
\centering
\begin{minipage}{.45\textwidth}
\centering
\includegraphics[scale=0.3]{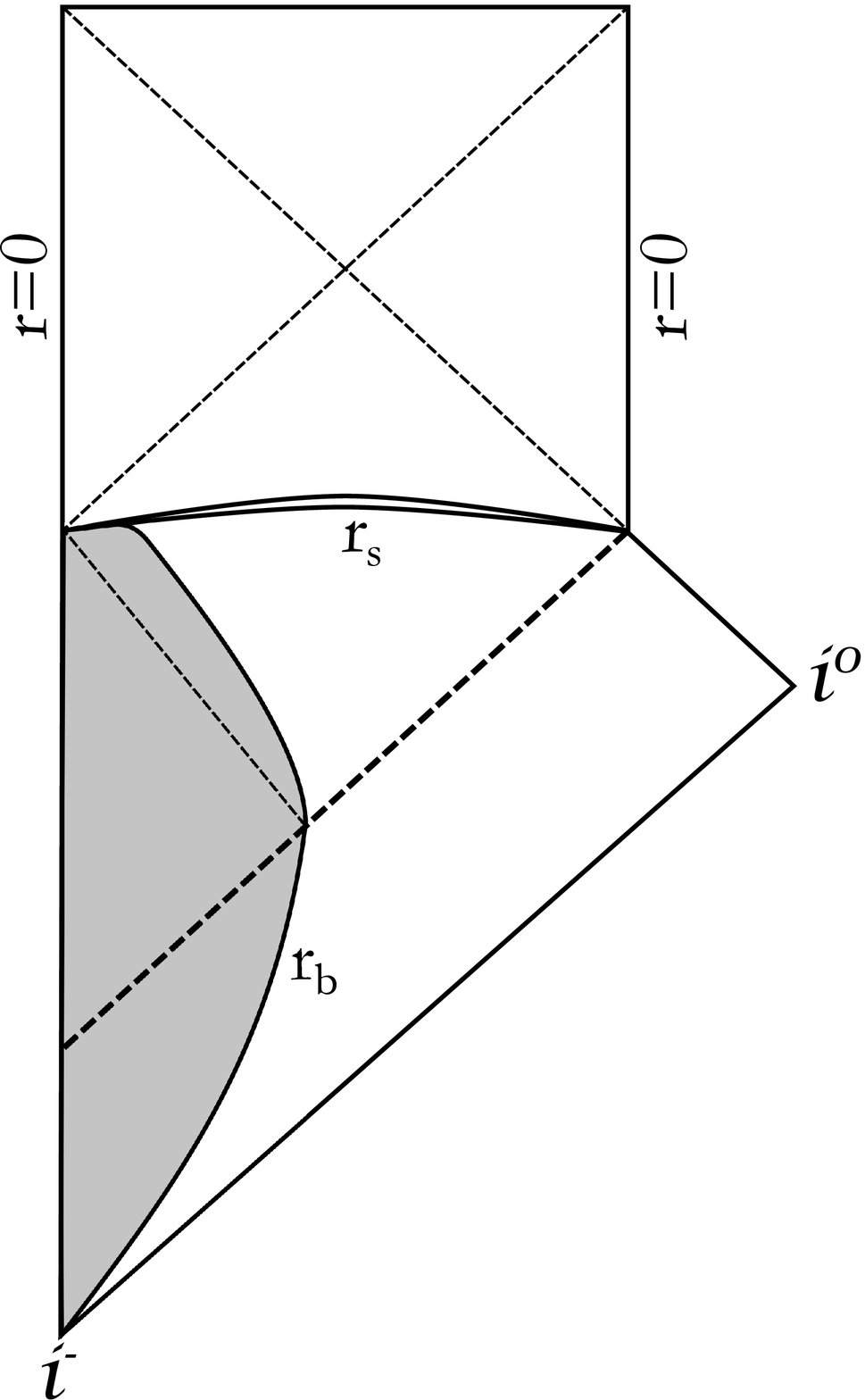}
\put(-45,110){I}
\put(-120,90){II}
\put(-89,200){III}
\end{minipage}
\hfill
\begin{minipage}{.45\textwidth}
\centering
\includegraphics[scale=0.3]{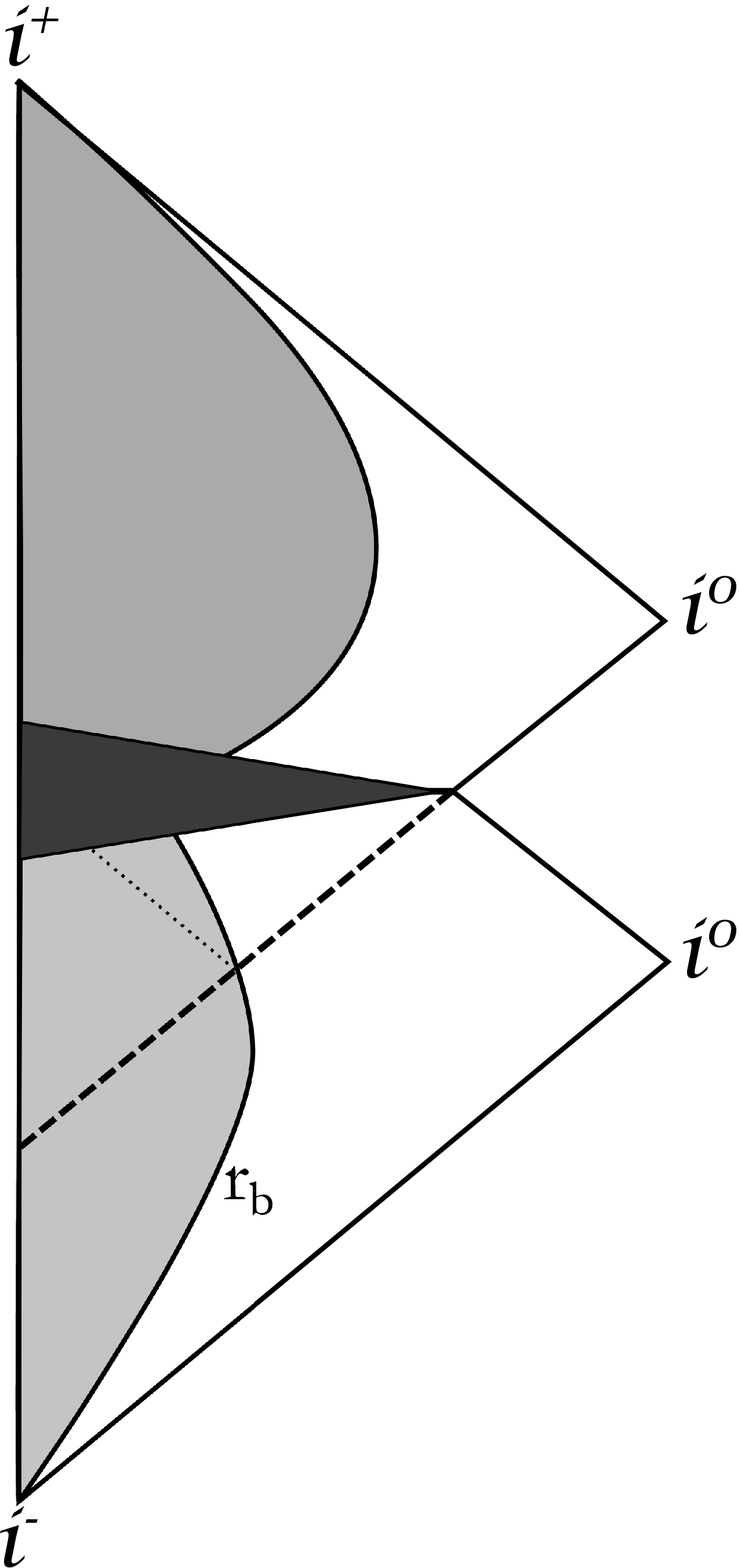}
\put(-45,88){I}
\put(-100,88){II}
\put(-100,170){III}
\end{minipage}
\caption{Two possible Penrose diagrams for the formation of a baby universe inside a black hole. Collapse of a matter cloud (grey region II within boundary $r_{\rm b}$) proceeds classically until after the formation of the event horizon (thick dashed line). The event horizon in not affected by quantum corrections close to the singularity. Far away observers (in region I) see the formation of a Schwarzschild black hole.
Left panel: The singularity at the end of collapse is removed and replaced by a junction surface (double solid line $r_{\rm s}$) to a DeSitter universe (region III). The thin dashed lines represent the Cauchy horizons (see \cite{baby}).
Right panel: When reaching the quantum gravity region (darker grey region) matter undergoes a phase transition and re-expands after the bounce (region III). However, as opposed to the white hole scenario, the expanding matter remains confined within the horizon and generates an expanding universe causally disconnected from the original one. The dynamics of horizons in the baby universe would depend on the properties of the expanding matter.}
\label{baby}
\end{figure}

This is related to an idea for the resolution of the Schwarzschild singularity that was developed in
\cite{baby}. 
In \cite{baby} it was shown that a matching through a transition layer could be performed between Schwarzschild and DeSitter in such a way that the singularity in the former is removed and a closed universe described by the latter develops inside the black hole.
The same idea can be extended to the case of black holes forming from collapse
(see left panel in figure \ref{baby}).
Since then the idea of universes evolving inside a black hole have been considered in a variety of contexts
(see for example \cite{pop} or \cite{hsu}). 
In
\cite{smolin}, 
it was argued that creation of universes inside black hole may provide an evolutionary mechanism that favours universes like the one we observe.
It is not surprising then that they appear also in solutions describing bouncing models from collapse.
For example, baby universes as emerging from the quantum bounce of a null shell in LQG have been suggested in
\cite{pul}.
Also, baby universes have been observed to appear for certain choices of parameters in Friedmann-Robertson-Walker models with quantum corrections coming from LQC
\cite{vid1}.
In the above models Hawking radiation is not considered.
However, universes inside a black hole may also provide clues towards the resolution of the information loss problem (see
\cite{smolin2}). 
In fact by allowing world-lines of particles to extend in the baby universe one avoids the loss of information that would happen at the singularity. In this case the Hawking radiation particles in the main universe would be entangled with the particles in the baby universe although they would exist in causally disconnected regions. The situation would resemble the information loss scenario for external observers up until the time in which the black hole horizon reaches Planckian size (which for astrophysical black holes is still longer than the current age of the universe). What would happen at that point is highly speculative and depends on whether predictability is preserved at that stage or not (see for example, \cite{HBU1} and \cite{HBU2}).

In order for the process that leads to the formation of a baby universe to happen, a phase transition is necessary for the collapsing matter at some point close to the bounce. A phase transition in the quantum regime could then be related to a topology change in the space-time thus allowing for the creation of a new asymptotically flat region for observers inside the black hole (see right panel in figure \ref{baby}).
The scenario can be made even more fascinating with the addition of a few speculations. Given the fact that the interior is causally disconnected from the exterior space-time it is entirely possible that the mass $M$ as measured by observers at infinity does not correspond to the mass $M$ for observers inside the baby universe. This could be a consequence of the failure of classical conservation laws across some surface (the horizon or phase transition surface, for example)
\cite{hell}. 
In fact quantum fluctuations and pair production near the bounce could effectively act as a `big bang' thus creating an entire universe like our own within the black hole.

\textbullet\ Periodic solutions: Finally, we shall briefly mention here another possibility that has been suggested in \cite{garay1} and \cite{garay2} but has not been thoroughly investigated so far. This is the case of a repeating cycle of collapses and bounces leading to a periodic solution. 
In the non dissipative case, with collapse starting with zero velocity from a finite radius, it would seem natural that periodic solutions may exist. The collapsing cloud would reach a minimum size and bounce to return to its original size with zero velocity again. Then the process would repeat unaltered. The black hole horizon would form and disappear at regular intervals, thus leading to a new kind of astrophysical phenomena. Unfortunately, due to the time delay of signals coming from the deep gravitational well of the pulsating object the intervals between two explosions may be too long to be experimentally observed.
In a more realistic scenario the oscillating object would lose part of its mass at every cycle and dissipative effects would dampen the oscillations making each subsequent `explosion' less powerful than the previous one. In this case, over long time scales, the object would slowly settle to a final static configuration described by a low mass, high density, exotic compact remnant (possibly without horizon).

In 
\cite{garay2}, \cite{bar3} and \cite{bar4} the possible existence of oscillating models was suggested. In \cite{garay1} a model with multiple bounces was proposed in which the object would lose energy after every bounce, via dissipation and possibly the emission of gravitational waves, to eventually settle to a `black star' final configuration. The geometry of these kind of oscillating solutions was discussed in \cite{barcelo-uni}.
However, it is important to stress that full analytical models describing oscillating scenarios are at present almost entirely unexplored.



\section{Remnants and phenomenology}\label{pheno}

When considering the modified collapse scenarios, the most important question for astrophysics is whether quantum gravity corrections to the black hole formation models have any observable effect in the real universe.
Astrophysical compact sources and cosmology are the only regions in the universe where one can hope to test such proposals. The issue has been debated for a long time (see for example,
\cite{liberati1} for proposed tests of quantum-gravity via cosmic rays or \cite{gupt} and references therein for possible observations of quantum-gravity effects in the early universe).
We have seen that it is possible that quantum corrected models for collapse induce modifications of the geometry by quantum effects in the near horizon region. This indicates that the consequences of a UV completion of gravity may be observable, at least in principle, also in a regime where classicality dominates.
Furthermore, this is not a far fetched speculation with no hope of any real tests being carried out. Precision measurement of astrophysical phenomena are improving every year and it is conceivable that soon we will have enough data from regions of space-time surrounding the locations where the horizon of black hole candidates should be. The presence or absence of non negligible quantum effect may then allow us to test the theories. In fact, it is indeed possible that some energetic phenomena already observed in the universe may have quantum gravitational origin, but their theoretical explanation has not been clarified yet.
Here we ask the question `what kind of phenomenology do these quantum corrected models entail'?
Can the outgoing matter coming out of the white hole horizon produce explosive phenomena (as suggested for example in 
\cite{dadhich})?
Or can the matter be radiated away over long times at low energy emission rates?
Is it possible that such phenomena have a distinctive gravitational wave signature?
Can a small, dense compact object of finite size survive the process? 
Or must all the matter be radiated away?
If a long lived horizon forms, can it be distinguished from the black hole event horizon?

\subsection{Compact objects}

Compact objects in GR have a long history, starting from the static and spherically symmetric constant density interior solution found by Karl Schwarzschild in 1916,
together with the more famous vacuum solution. Later Tolman considered several classes of interior solutions
\cite{tolman},
one of which was used by Oppenheimer and Volkoff to study an equilibrium state for matter that describes a cold Fermi gas and could be used to model a neutron star
\cite{OV}.
The Oppenheimer-Volkoff model is the general relativistic equivalent of the model that was developed by Chandrasekhar
using special relativistic effects and electron degeneracy pressure to show that there exist an upper mass limit for white dwarfs
\cite{chandra}.
A key ingredient of the relativistic theory of stellar structure is the Tolman-Oppenheimer-Volkoff equation which constraints the structure of the compact object in order to preserve it in equilibrium. The specific form of the Tolman-Oppenheimer-Volkoff equation depends on the equation of state describing the matter model. Studying the mass-radius relation coming from this constraint provides the upper bound on the mass of the compact object. For a given equation of state, objects above a certain limit can not stay in equilibrium and must collapse. 
Within classical GR it follows that there is no way to build a stable object above the neutron degeneracy pressure and therefore black holes are the only possible outcome once that threshold is passed.

One intriguing alternative that is suggested by the bouncing scenarios described above is that exotic compact objects as leftovers from collapse may exist in nature.
This can already be seen by the behaviour of the apparent horizon, which shows a threshold (with scale determined by the parameter $\rho_{\rm cr}$) below which no trapped surfaced form. This means that the collapse and bounce of an object small enough would never form any horizon. Then one naturally wonders whether a stable object may exist below this threshold.
In order to understand whether such possibility could be realized in the universe one has to understand the phenomenology of the collapse scenarios. What happens to matter once it undergoes a bounce in a regime where quantum gravity prevails? Is it possible that an island of stability exists for matter at densities above the neutron degeneracy threshold?
Indeed from all of the discussion above it seems that asking the question whether the observed astrophysical black hole candidates are indeed black holes in the usual sense
is not a mere theoretical exercise
\cite{compact-visser}.

As we have seen, in some models, the region where quantum gravity effects are significant may extend until the horizon and beyond. Also we have seen that it is possible for quantum effects to destroy the horizon. Modified Schwarzschild solutions, such as the ones describing regular black holes, and solutions describing horizonless exotic objects seem to be perfectly acceptable. Remember that depending on the value of the critical mass a solution for a regular static black hole like the one in \cite{hayward} can have multiple horizons or none, which implies that in the horizonless case the compact remnant would be visible to far away observers.

In recent times different kinds of models describing `exotic' interiors have been proposed. These may be in the form of remnants that are left after Hawking evaporation or some other process destroys the black hole horizon, or compact objects of finite size larger than the Schwarzschild radius whose formation prevents the formation of the horizon. These objects can be roughly divided into four categories: (i) Planck size remnants left after evaporation, (ii) Planck size remnants left after some other mechanism destroys the horizon (like for example, Planck stars), (iii) finite size horizonless objects with radius greater than the horizon (like for example, quark stars and boson stars) and (iv) finite size horizonless objects with radius slightly larger than the horizon (like for example, gravastars and black stars). It is important to stress that the dynamical evolutions that lead to the formation of each of these types of objects vary greatly from model to model and this implies that their observational signatures are going to be very different.

\textbullet\ 
When considering horizonless compact objects, gravastars (for `{\em gra}vitational {\em va}cuum {\em stars}') constitute an alternative to black holes as the final product of gravitational collapse. They were first suggested in
\cite{grava1}.
The main idea is that just outside the Schwarzschild radius, matter undergoes a phase transition and `condensates' at a finite radius slightly larger than the horizon radius, thus leaving an horizonless compact object composed by a shell with stiff equation of state $p=\rho$ that separates the Schwarzschild exterior from a DeSitter interior. 
The existence of a physical surface outside the Schwarzschild radius, as opposed to the event horizon, is the main indication that such models could be tested by observations
\cite{grava1b}.
In this respect gravastars constitute a valid alternative to black holes, they are not plagued by many of the shortcomings of classical black holes and they naturally bridge the classical regime with the quantum regime.
In order to establish whether such proposed models can exist in the universe one needs to understand their physical and observational properties.
Stability of gravastars was studied in
\cite{grava2}, while in
\cite{grava3}
it was shown that `realistic' gravastars must necessarily
have anisotropic pressures.
Also the spectrum of quasinormal modes of such objects is considerably different
from that of a black hole thus suggesting that it may be possible to distinguish them observationally
\cite{grava4}.
However, in 
\cite{grava5}
it was suggested that, despite the difference in quasinormal modes, gravastar mergers could produce a gravitational wave signature (in particularly a ringdown phase, which is usually associated with the existence of the horizon) similar to that of black hole mergers. Therefore great caution should be used when claiming that the ringdown of detected gravitational wave signals is a direct proof of the existence of the black hole horizon.
Finally we should note that in
\cite{grava6}
it was shown that gravastars could be distinguished from black holes also via direct imaging of their shadow.

\textbullet\ More recently, compact remnants, as alternative to black holes, resulting from semi-classical collapse were suggested, with the name of `black stars', by Barcel\'o et al. in 
\cite{bar1}. These are compact objects filled with matter, and with an actual surface, that are denser than neutron stars. Black stars are supported in equilibrium by the pressure provided by quantum vacuum polarization and can be understood as the limiting case of an isothermal sphere, where every shell of the black star is close to the horizon limit. Due to their compactness black stars have large redshift that can observationally mimic black holes. Furthermore, in \cite{blackstar} it was shown that such objects can emit Hawking-like radiation similarly to black holes.

\textbullet\  
Remnants have also been considered as possible leftovers from complete evaporation of black holes
(see\cite{hayward}). Such objects arise naturally when considering Planck scale cut offs to Hawking evaporation and have sizes (of the order of Planck scales) that depend on the information that they contain. These models may offer a solution to the information loss problem (see for example 
\cite{hus4} and \cite{gid1}). 
Also, quantum effects on inhomogenous dust collapse were considered in order to construct a solution describing a compact remnant in place of a naked singularity in
\cite{vaz}.

\begin{figure}[ht]
\centering
\begin{minipage}{.45\textwidth}
\centering
\includegraphics[scale=0.3]{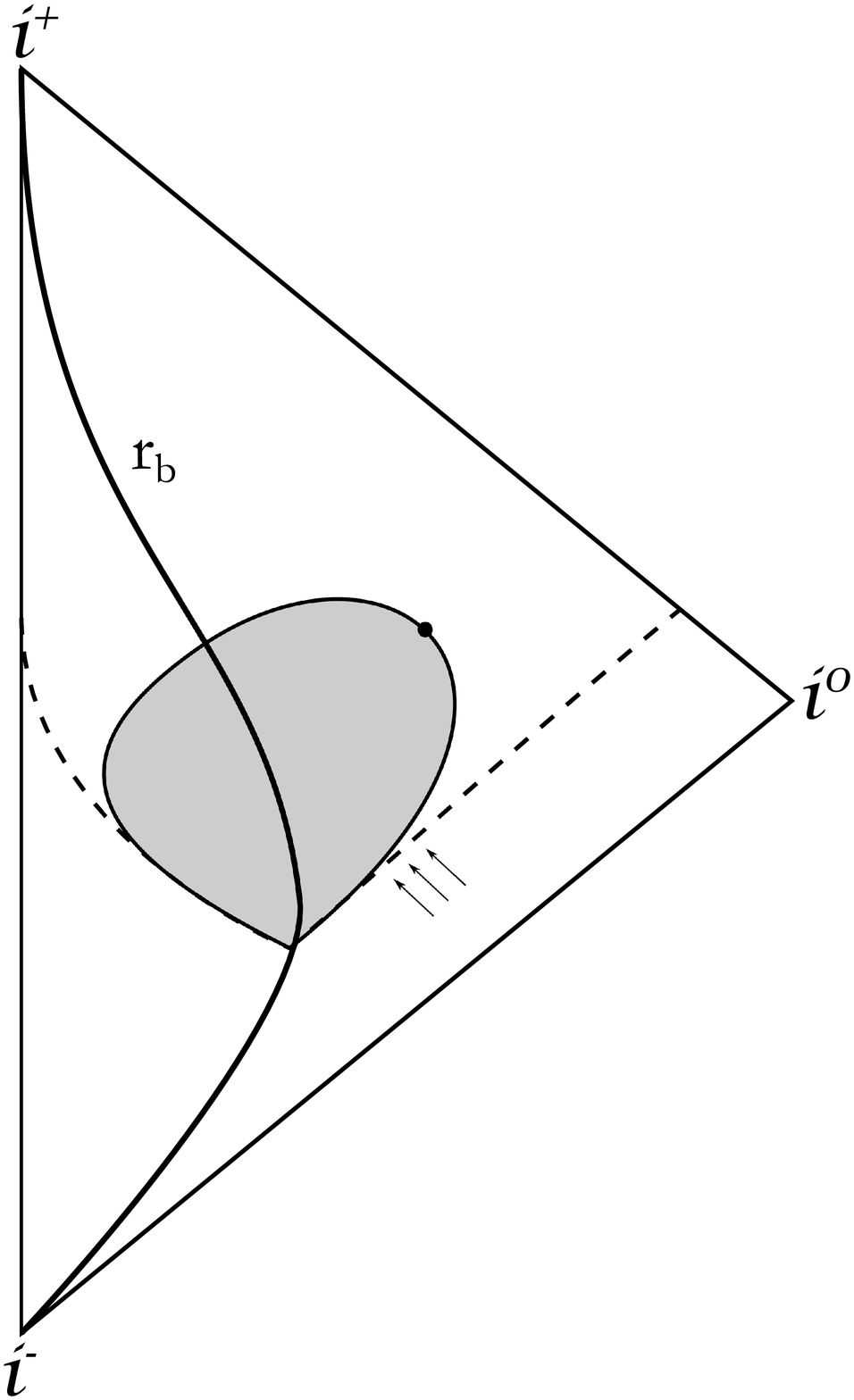}
\put(-73,140){P}
\put(-70,130){oh}
\put(-92,148){ih}
\put(-130,119){ah}
\end{minipage}
\hfill
\begin{minipage}{.45\textwidth}
\centering
\includegraphics[scale=0.3]{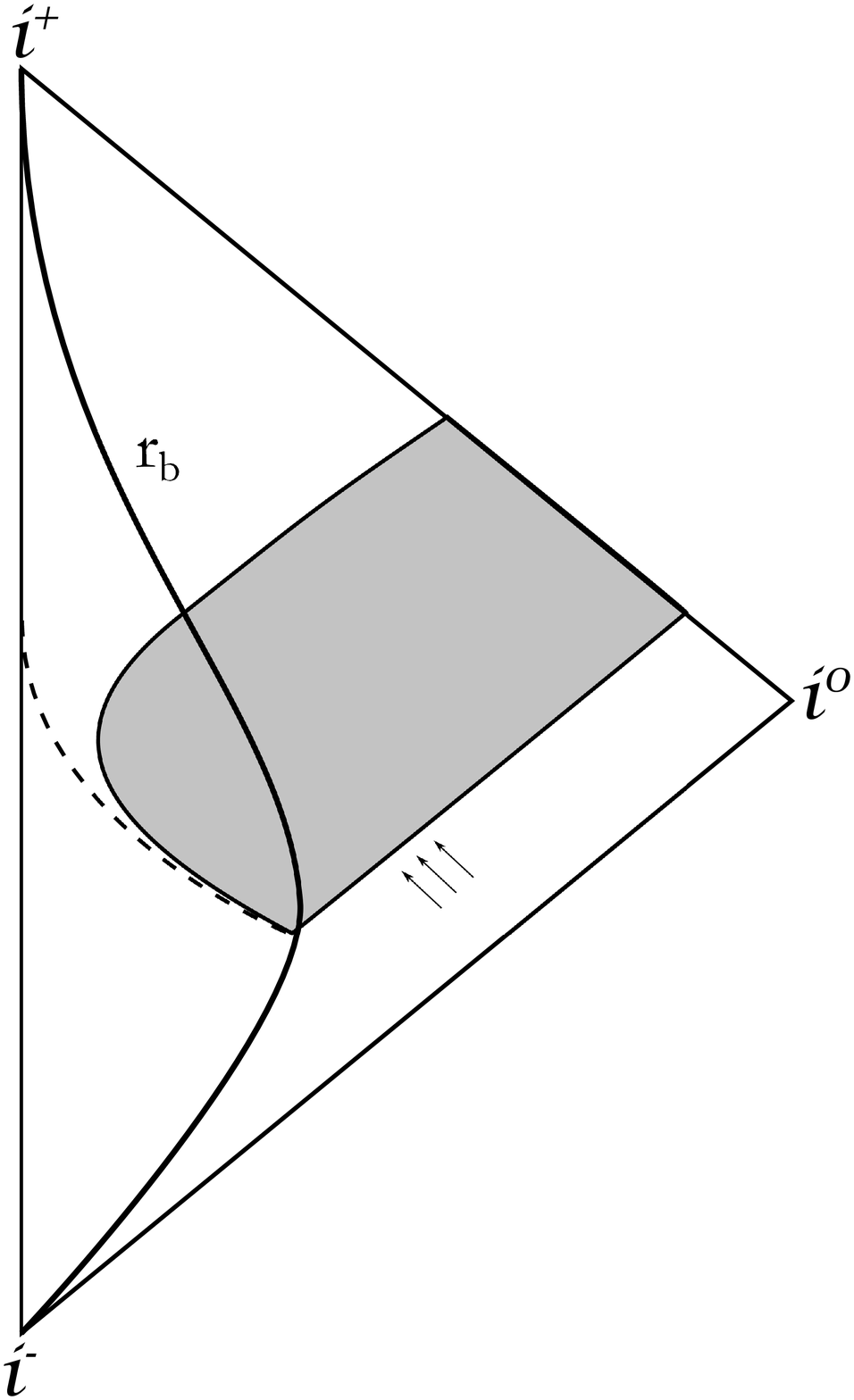}
\put(-64,110){oh}
\put(-100,166){ih}
\put(-131,120){ah}
\end{minipage}
\caption{Penrose diagrams of the formation of exotic compact objects from collapse with semi-classical corrections. The thick solid line describes the boundary of the collapsing body $r_b$. Dashed lines describe apparent horizon (curved) and event horizon (straight) in the classical case. 
Left panel: Quantum effects modify the geometry in the exterior and the outer horizon (oh) shrinks. The outer horizon eventually annihilates with the expanding inner horizon (ih) at the point $P$, the trapped region (grey area) lives for a finite time. The geometry tends towards the classical case at large distances, while for large $t$ near the compact object (up to a radius greater than the Schwarzschild radius) there are no trapped regions. 
Right panel: The outer horizon (oh) is well described by the black hole event horizon. Quantum effects result in the formation of an inner horizon (ih) and the system settles to a regular black hole geometry. The trapped region (grey area) is enclosed within the two horizons in the exterior and the apparent horizon (ah) in the interior (solid lines).}
\label{remnant}
\end{figure}

\textbullet\ As said, remnants are Planck size objects that are left after the disappearance of the horizon. Among the various kinds of exotic remnants that have been suggested in the literature, one that is directly related to the bouncing collapse models is the so called `Planck star'. It was proposed in 
\cite{ps1} as the ultra-compact remnant of a massive star whose collapse has been halted by repulsive quantum effects. As in the models seen above, the limit at which quantum effects become important is given by the Planck density, which implies a final size for the object much larger than the Planck length. For example the model discussed in \cite{ps2} has a size of the order of $10^{-10}$cm, which, while being smaller than the size of an hydrogen atom is still much larger that the Planck scale. In this scenario the quantum matter distribution is confined behind an inner and an outer horizon (see right panel of figure \ref{remnant}) and the Planck star has a lifetime, as seen by far away observers, of the order of $M^3$, comparable to the Hawking evaporation time. In 
\cite{ps2}
the effective metric describing a Planck star was modelled as a modification of a regular black hole metric (such as the one discussed in \cite{hayward}).
In \cite{ps3}
the authors considered the possibility of detecting the final explosion of a primordial Planck star. The main idea is that if such objects formed in the early universe, then, given the average lifetime of the horizon it is possible that the disappearance of their horizon is occurring today. If this process is explosive and accompanied by the emission of gamma rays, then there is some non zero chance of detecting it with modern telescopes (see also \cite{rov-new}).


\textbullet\ Given the existence of several proposals for exotic compact objects, it is interesting to investigate under what conditions gravitational collapse halts at a finite or zero radius, without the formation of a singularity and without a bounce. Is it possible to construct a dynamical evolution that does not bounce and for which collapse stops as $t$ goes to infinity? To answer this question, consider for simplicity the metric \eqref{interior} for a homogeneous fluid, whose dynamics is described by the evolution of the scale factor $a(t)$. Then the condition for the formation of a compact remnant is
\be \label{eq}
\dot{a}=\ddot{a}=0  \; .
\ee
In the classical case we get
\bea
\dot{a}&=&-\sqrt{\frac{M}{a}+b_0}  \; , \\ \label{addot-cl}
\ddot{a}&=&\frac{1}{2}\left(\frac{M_{,a}}{a}-\frac{M}{a^2}\right)=-\frac{a}{2}\left(p+\frac{\rho}{3}\right)  \; .
\eea
From the above we see that if $M\neq 0$ collapse can halt only in the bound case where $b_0<0$. In the case where $M$ goes to zero then all matter is radiated away and collapse stops leaving behind Minkowski.
Also from equation \eqref{addot-cl} we see that for collapse to stop the equation of state must tend to $p=-\rho/3$.
In the semi-classical case however we have
\bea\label{adot}
\dot{a}&=&-\sqrt{\frac{M}{a}\left(1-\frac{3M}{\rho_{\rm cr}a^3}\right)+b_0}  \; , \\ \label{addot}
\ddot{a}&=&\frac{1}{2}\left[\frac{M_{,a}}{a}\left(1-\frac{6M}{\rho_{\rm cr}a^3}\right)-\frac{M}{a^2}\left(1-\frac{12M}{\rho_{\rm cr}a^3}\right)\right]  \; ,
\eea
from which we see that in the marginally bound case ($b_0=0$) as $a\rightarrow a_{\rm cr}$ (with $a_{\rm cr}^3=3M/\rho_{\rm cr}$) we get $\dot{a}\rightarrow 0$, which is the condition for occurrence of the bounce. Also in this case
\be \label{eq1}
\ddot{a}\rightarrow \frac{a_{\rm cr}}{2}\left(\rho_{\rm cr}-\frac{M_{,a}}{a_{\rm cr}^2}\right)  \; .
\ee 
Now equation \eqref{p} tells us that the second term in equation \eqref{eq1} is nothing but the pressure. Therefore a fluid that tends to an equation of state of the type $p=-\rho$ (cosmological constant) in the limit of $\rho$ going to $\rho_{\rm cr}$ will halt collapse asymptotically at a finite or zero value for the scale factor.
In the case of $a_{\rm cr}=0$ either the whole mass of the initial configuration is radiated away or the singularity occurs as $t$ goes to infinity. 
In the case of $a_{\rm cr}>0$ 
the apparent horizon curve $r_{\rm ah}(t)$ in the interior follows again equation \eqref{ah}, with the effective mass in place of the physical mass and the trapped region develops at the boundary in the regime where classical behaviour holds. In the exterior an horizon that follows the event horizon appears at the same time. As the density approaches the characteristic values of the critical scale, the system enters the quantum corrected regime and the apparent horizon curve deviates from the classical trajectory after reaching a minimum value. Then $r_{\rm ah}(t)$ moves outwards to cross the boundary at a certain time, where it connects with the second (inner) horizon of the exterior metric. If the quantum effects reach the exterior to cause the outer horizon to shrink then the two horizons in the exterior meet and annihilate leaving the compact remnant visible to far away observers 
(see left panel of figure \ref{remnant}).
If the exterior is described by a regular black hole solution then the two horizons live for a long time (of the order of the Hawking evaporation time) and the space-time is described by a Planck star solution with dark energy equation of state
(see right panel in figure \ref{remnant}).

\textbullet\ We have seen that if we wish for collapse of an homogeneous perfect fluid to halt asymptotically we need to impose an equation of state that tends to a dark energy fluid ($p=\lambda\rho$ with $\lambda<-1/3$) in the limit of densities reaching the maximum density. In this case collapse can stop to produce a final `dark energy like' equilibrium configuration. This is closely related to the gravastar model since the dark energy equation of state for the fluid's interior is reminiscent of the DeSitter core of gravastars. 
Compact objects with a dark energy equation of state have been considered in
\cite{ds1}
as possible extensions of gravastar models.
In
\cite{ds2}
a varying dark energy equation of state $p=\lambda\rho$ with $\lambda<-1/3$ was considered. If the parameter $\lambda$ is allowed to cross the `phantom' threshold $\lambda=-1$ then a topology change may occur and the repulsive effects can be used to construct stable wormhole solutions.
Similarly to the case of gravastars, anisotropies may turn out to be important for the construction of stable dark energy stars. For example, in
\cite{ds4}
one of such anisotropic models was studied in which the star's interior structure is comprised of two fluids, one baryonic fluid and one repulsive dark energy fluid.

\subsection{A toy model of collapse to a dark energy star}

\begin{figure}[h]
\centering
\begin{minipage}{.45\textwidth}
\centering
\includegraphics[scale=0.4]{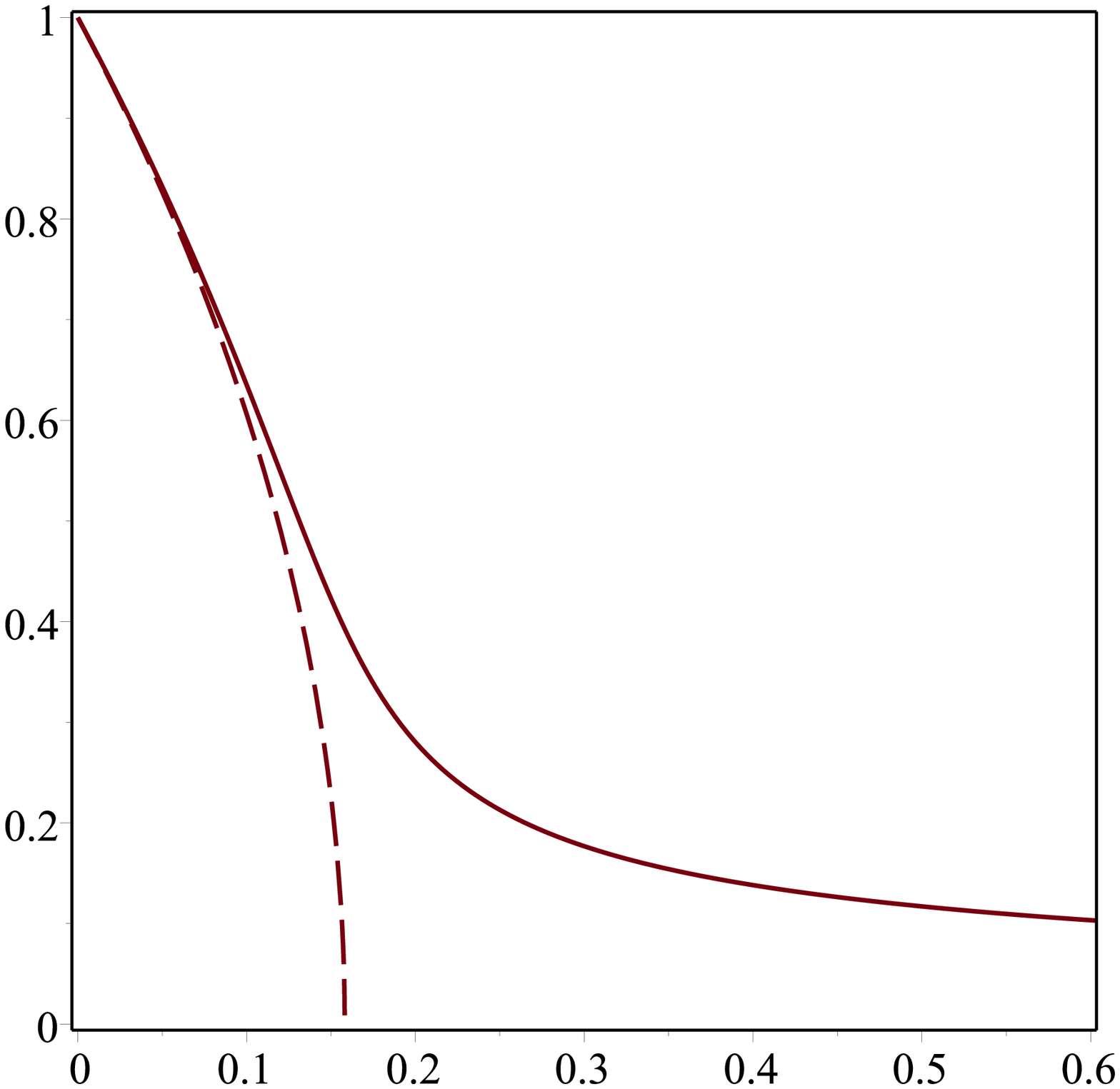}
\put(-207,200){$a$}
\put(-25,15){$t$}
\end{minipage}
\hfill
\begin{minipage}{.45\textwidth}
\centering
\includegraphics[scale=0.4]{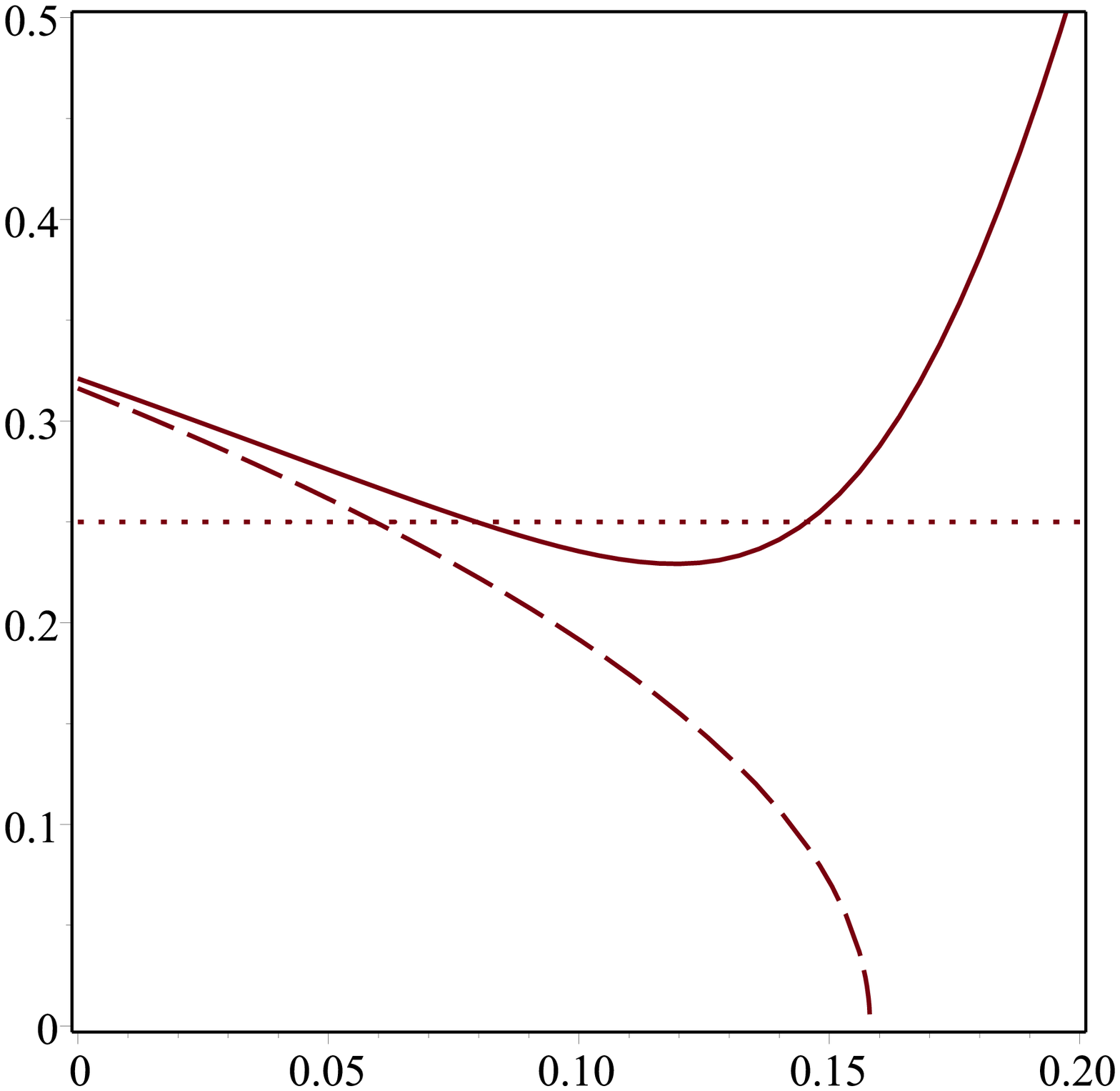}
\put(-207,200){$r$}
\put(-25,15){$t$}
\put(-212,117){$r_{b}$}
\end{minipage}
\caption{Comparison between classical homogeneous dust collapse (dashed lines) and semi-classical collapse leading to an almost static compact remnant (solid lines) for a fluid approaching a `dark energy' equation of state. 
Left panel: Comparison of the scale factors $a(t)$. In the classical case $a$ goes to zero in a finite time. In the semi-classical case $a$ goes to zero as $t$ goes to infinity.
Right panel: Comparison of the apparent horizon curves $r_{\rm ah}(t)$. In the classical case $r_{\rm ah}$ goes to zero in a finite time. In the semi-classical case the apparent horizon reaches a minimum value and then tends to infinity as $t$ grows, thus indicating that eventually it crosses the boundary of the collapsing object (indicated here by the dotted line of the co-moving radius $r_b$). 
The plots have values of the parameters chosen as follows: $M_0=10$, $\lambda=1/3$, $\rho_{\rm cr}=1000$, $r_b=0.25$. Note that in order to have the initial boundary surface not trapped one has to consider $r_b<r_{\rm ah}(0)$ so that if $r_b$ is smaller than the minimum value of $r_{\rm ah}$ no trapped surfaces form.}
\label{toy}
\end{figure}

In this final part of this section we wish to discuss a new dynamical toy model for collapse that leads to the formation of one of such objects.
The easiest way to construct a toy model that settles to a finite sized compact remnant is to make an `ad hoc' choice for the equation of state in such a manner that the conditions for equilibrium \eqref{eq} are met. This means, for example, considering an equation of state that is linear in the weak field and tends to a dark energy fluid in the strong field (as the density approaches the critical value). The physical interpretation of such a choice will be left aside for the moment.
 The simplest way of implementing this is to just assume a dark energy fluid equation of state in the form $p=-\rho$ to begin with. Then the equation of motion for both the classical and semi-classical scenarios gives the scale factor as $a(t)=e^{-t/T_{\rm cr}}$, where the time-scale parameter $T_{\rm cr}$ is given by $T_{\rm cr}^{-1}=\sqrt{M_0}$ in the classical case and $T_{\rm cr}^{-1}=\sqrt{M_0(1-3M_0/\rho_{\rm cr})}$ in the semi-classical case.
 From this we see that $a$ goes to zero as $t$ goes to infinity and therefore no compact remnant of finite size is left over from collapse asymptotically. However, if the exterior horizon were to disappear due to quantum effects such as the ones described in section \ref{open}, this would imply the existence of a long lived evaporating object. We can guess that if the dark energy equation of state dominates in the strong field a similar behaviour will appear in general.
 To check this we consider a toy model where we add a suitable quadratic correction in $\rho$ to the linear equation of state.
 Therefore we consider
 \be\label{eos}
 p=\lambda\rho-\frac{1+\lambda}{\rho_{\rm cr}}\rho^2  \; .
 \ee
Note that for positive $\lambda$ the pressure reaches the maximum value $p_{\rm max}=\lambda^2\rho_{\rm cr}/(4+4\lambda)$ at densities below the critical scale, while for $\rho>\rho_{\rm cr}(3\lambda+1)/(3\lambda+3)$ the equation of state crosses into the dark energy regime.
Then by using equation \eqref{eos} and substituting $p$ and $\rho$ from equations \eqref{p} and \eqref{rho} we get
 \be
 \frac{dM}{da}=\frac{3M}{a}\left(\frac{1+\lambda}{\rho_{\rm cr}}\frac{3M}{a^3}-\lambda\right)  \; ,
 \ee
 which can be easily solved to give
 \be
 M(a)= \frac{M_0}{a^{3\lambda}+\frac{3M_0}{\rho_{\rm cr}a^3}}  \; .
 \ee
 Note that for $\rho_{\rm cr}\rightarrow +\infty$ we recover the usual perfect fluid with linear equation of state. Also note that as $a\rightarrow a_{\rm cr}$ we get $M_{,a}\rightarrow a_{\rm cr}^2\rho_{\rm cr}$, which, once plugged into \eqref{eq}, gives $\ddot{a}\rightarrow 0$.
 By plugging $M$ into the inverse of the equation of motion \eqref{adot} with initial condition $t(1)=0$ we can integrate to get $t(a)$ (note that to be able to invert we must verify that $a$ is monotonic, which is true until the time at which $\dot{a}=0$) and therefore the scale factor $a(t)$ after inverting. The evolution of the scale factor for this model is shown in the left panel of figure \ref{toy}.
The evolution of the apparent horizon follows the classical behaviour at early times while as $t$ grows $r_{\rm ah}$ goes to infinity. This suggests that the apparent horizon will cross the boundary at a finite co-moving time, thus developing a situation similar to the one discussed in the previous section (see right panel in figure \ref{toy}).
In the case where the exterior metric is modified by quantum effects outside the event horizon we retrieve a Planck density compact remnant that slowly `evaporates'.
 
\subsection{Future observations}

The bouncing scenarios described in the previous sections present some interesting theoretical challenges, however the most important question is whether they are related to any phenomenon occurring in the real universe. If the UV corrections in the strong field were confined to a close neighborhood of the center with a classical event horizon developing in the exterior, then we would have no hope of testing these proposals experimentally. However, as we have seen, the classical black hole geometry can be significantly affected by quantum correction even in the weak gravity regime and outside the horizon. Therefore it is legitimate to consider the observational signatures that these models may bear on astrophysical phenomena.
To do so, we need to understand the phenomenological features of such models. Then we can hope to check their validity against astrophysical observations.
Astrophysical data of black hole candidates today comes essentially from four types of observations: Spectrum from accretion disks, direct imaging of the black hole shadow (soon to be possible with very long baseline interferometry such as the Event Horizon Telescope), gravitational lensing and gravitational waves from inspirals of compact objects.
In fact, some of these models may already be constrained by observations, at least in principle. For example, short lived black holes that end their lives in a powerful explosions should already be detectable either via optical or gravitational wave signals, and the probability of observing one is related to their statistical occurrence in the universe.
To summarize there are two possible classes of deviations from the classical scenario that can in principle be tested: (i) static and quasi-static final configurations, like regular black holes or horizonless exotic compact objects and (ii) dynamical scenarios related to the disappearance of the horizon, which could imply gamma rays explosions or the emission of matter from the white hole phase.
In the first case one can hope to observe and measure the optical properties of accretion disks in the vicinity of the source. For example, soon we might be able to observe whether deviations from the classical Kerr geometry occur in the vicinity of the super-massive black hole candidate at the center of our galaxy.
In the second case one can hope to observe optical phenomena like gamma rays coming from the black hole to white hole transition or the emission of gravitational waves from the bouncing phase. Gravitational waves may also provide indications of deviations from the classical black hole geometry in mergers of binary systems composed of alternative objects.

\textbullet\ Geodesics and lensing:
In
\cite{gid3}
it was argued that modification to the near horizon geometry of the black hole due to the effective quantum corrections could sufficiently alter geodesics in the space-time thus offering a concrete possibility to test quantitatively the form of the modifications. 
Accretion disks around regular black holes and horizonless space-times were considered for example in
\cite{stu}. 
The authors showed that when an horizon is present the differences between the modified geometry and the classical black hole geometry may be impossible to be determined efficiently from observations. On the other hand, regular space-times without horizon are clearly distinguishable, even qualitatively, from black holes.
Gravitational lensing by a regular black hole was studied in
\cite{eiroa},
while in
\cite{jap}
the authors considered geodesic motion for massive particles and photons in the Hayward metric given in
\cite{hayward}
and showed that by observing the shadow of a black hole candidate it may be possible to distinguish whether the space-time is described by this metric or the Schwarzschild metric.
Further in 
\cite{me-cosimo}
the k$\alpha$ iron line from accretion disks around exotic compact objects was simulated to show that such disks may produce a spectrum considerably different from the black hole case. In particular, if one allows for circular geodesics to extend all the way to the center and no horizon to be present, then the drop at high frequencies of the spectral luminosity distribution that signals the existence of the event horizon disappears, replaced by a more gradual power law decrease.

\textbullet\
Direct imaging: In the above category, while at the same time standing on its own, is the case of the super-massive black hole at the center of the Milky Way, Sagittarius A* (Sgr A*). The main reason is that, due to its relative vicinity, the near horizon region may be directly observable in the case of Sgr A*. The black hole candidate (as well as the one at the center of the galaxy M87) is massive enough and close enough to have an angular diameter in the sky that may be resolved by Very Large Baseline Interferometry observatories thus offering the chance to obtain direct imaging of the black hole shadow \cite{vlbi}. 
For a review of possible observations of SGR A* see
\cite{Goddi:2016jrs}.

Possible deviations from the Kerr geometry in the vicinity of Sgr A* have been considered by many authors
(see for example \cite{johannsen} and \cite{bambi} and references therein).
In 
\cite{haggard}
it was suggested that while quantum corrections outside the horizon may be small, they may have cumulative effects over time and thus the classical behaviour near the horizon may be lost after a sufficiently long time.
In this respect we should add here that, while it is often stated that quantum effects become important for scales of the order of the Planck length (of the order of $10^{-33}$cm), it is more plausible that it is the Planck density that sets the threshold. Therefore for a stellar mass object the critical length scale at which quantum effects become important would be of the order of $10^{-20}$cm, thus much larger than the Planck length, even though still much smaller than the horizon scale. 
Also, this argument is based on the conservative assumption that no other repulsive effects occur between the scale related to the Planck density scale and the classical energy scale.

\textbullet\ Observing white holes:
If the bounce turns the black hole into a white hole, then how does the expanding matter emerge from the horizon?
Roughly speaking there are two main possibilities: An explosive event, where most of the energy is released over a very short time, or an `evaporation' type of process in which the energy is released slowly over a long time. Either way the event should be associated with some kind of electromagnetic emission.
Electromagnetic signals from primordial black holes exploding into white holes were considered in
\cite{vidotto} and \cite{rov-new},
while in
\cite{frb}
a possible connection of such models with observed Fast Radio Bursts was suggested.
It should be noted that if such phenomena occur in the universe then, in order to be detected observationally, there must be nothing else in the neighborhood of the bouncing object. If the white hole is surrounded by a gas cloud or by the remnant of the exploding progenitor star, then the radiation coming from the white hole region will undergo several processes before reaching the observer, thus possibly destroying the quantum signature of the signal.
Models for Gamma Ray Bursts explosions, for example, consider a region for the emission of gamma rays that is far from the core of the object. This is a general feature also in stellar collapse. The light coming from supernovae explosions is also produced far from the core of the collapsing star. Therefore the connection between the processes happening at the core, such as the bouncing scenarios described here, and observable effects is far from trivial. In a similar way, if matter is present outside the bouncing core, it could be that several processes occur after the outgoing matter passes the white hole horizon so that the light reaching far away observers would bear almost no information regarding what happens in the vicinity of the white hole.

\textbullet\ Gravitational wave signatures: Another method that may be soon available to test these hypoteheses is the observation of gravitational waves.
The waveforms, from inspiral of black hole candidates observed by LIGO
\cite{ligo}, 
are in perfect agreement with the predictions of GR. However it is possible that modifications to the black hole geometry outside the horizon may cause some yet unseen deviations of the gravitational wave signals from the classical predictions of GR. 
These deviations would be related to the length scale at which modifications extend outside the horizon, so that, in principle, current observations could already exclude models for which large scale deviations would produce a significantly different waveform
(see for example \cite{gid4}).
However in 
\cite{Konoplya:2016pmh}
it was shown that, due to the large uncertainty in the determination of mass and angular momentum of the merging black holes, the present observed gravitational wave signals do not rule out alternative theories of gravity. The possibility that such signals may be produced by merging wormholes was investigated in
\cite{Konoplya:2016hmd}
while in
\cite{Konoplya:2011qq}
the authors reviewed quasinormal modes in merger of black holes in different theories.

Another possibility is given by gravitational waves emitted by matter emerging from the horizon during the white hole phase. This is, in some sense, the white hole equivalent of gravitational waves produced during collapse to a black hole.
Then, the discovery of a distinctive gravitational wave signature from such exploding phenomena could help to determine what kind of scenario (if any) occurs in nature.
In this context, there are two kinds of instabilities that may play important roles in the observational outcome of bouncing scenarios. The white hole instability, that was described in section \ref{trans}, and the ergoregion instability, which states that rapidly rotating objects with an ergoregion but no horizon are unstable. This instability is a result of scattered waves that have larger amplitudes than incident waves, thus leading to an energy loss by the scattering body that may grow exponentially.
The instability occurs in any rotating object without an horizon but becomes non negligible only in very compact objects with high angular momentum
(see \cite{zeld1}, \cite{staro} and \cite{beck}).
The white hole instability suggests that the white hole phase must be short compared to the black hole phase, and therefore most of the matter from the outgoing cloud must be released in a short time.
The ergoregion instability suggests that as gravitational waves are produced by a rotating cloud coming out from the horizon the emission must also happen over short timescales, and therefore the amount of energy liberated per unit time must be large. Both arguments point to an explosive phenomenon for the phenomenological description of the outgoing matter emerging of the white hole horizon.



\section{Discussion}\label{discussion}

The final fate of gravitational collapse has been a very fruitful area of research in GR for decades. Its importance spans from fundamental questions about the nature of gravity to astrophysics.

\textbullet\ 
From the theoretical standpoint, black holes sit at the crossroad of several different disciplines. It is clear today that the black hole horizon is not a purely classical entity and it requires the addition of quantum physics to be properly understood. The description of the horizon that comes from our present understanding of GR and quantum mechanics is incomplete and presents some puzzles that, once solved, may open the way to a theory of quantum gravity. 
The final fate of collapse and the true nature of black holes have important consequences for long standing open questions related to information loss, unitarity and black hole firewalls
(for recent reviews see \cite{info}, \cite{info2}, \cite{info3} and references therein).
In the present article we have chosen to avoid a detailed discussion of such problems in order to concentrate on the astrophysical aspects. Nevertheless it is important to remark that both issues are two sides of the same coin and that a change in our understanding of one side will necessarily lead to a change on the other side.
For example, we have seen how a well defined UV completion of GR may lead to the resolution of the classical black hole singularity and this may eventually lead to the solution of the problems mentioned above.

\textbullet\
From the point of view of astrophysics, the quantum resolution of classical singularities that arise in dynamical models at the end of collapse, means that black holes, as described by the Schwarzschild and Kerr solutions, may not exist in the universe. Astrophysical black holes may posses horizons with a finite lifetime and may turn into white holes towards the end of their lives, possibly leaving behind some remnant of intrinsically quantum nature. 
This leads naturally to the question whether such predictions can be tested against observations. We have seen that some modifications to the dynamical behaviour of the collapsing cloud in the strong field may bear implications for the space-time geometry also at distances where one would expect only classical solutions to hold. Such models predict a change in the near horizon geometry that may have observable consequences. The effects of such change vary greatly from model to model and there is hope that present or future observations will be able to put constraints on the proposed scenarios.

\textbullet\ All the models studied in the literature are extremely simplified.
In particularly marginally bound collapse of adiabatic, non dissipative, fluids, leads to the complete time reversal of the solution after the bounce.
The physical validity of these solutions relies on the assumption that more realistic models will not change the qualitative picture. Therefore one future direction to explore implies considering improved theoretical models and it is safe to assume that numerical simulations will be necessary in order to solve more realistic scenarios.
Another question that future research will address is the possibility of the existence of periodic solutions. Such `oscillating' black holes, periodically drifting inside and outside the horizon, could hint at new astrophysical phenomena.

\textbullet\ As already mentioned, future observational tests of gravity in the strong field will put constraints on these proposals and hopefully (if there is a quantum signature in the near horizon geometry) indicate the way forward. This is not a far fetched speculation. Measurement of the properties of the gravitational field near the horizon of black hole candidates are already possible with gravitational wave detectors such as LIGO and VIRGO and at present it can not be excluded that they posses a signature of quantum gravitational nature.
Gravitational wave signals from exotic compact objects have been discussed in 
\cite{gw1}, while in
\cite{gw2} `echoes' in the signals arising from deviations from classical GR have been discussed.
Macroscopic effects in the signals due to corrections in the near-horizon geometry have been used in
\cite{gw3} in order to classify different proposals that present echoes and distinguish them from models that do not present echoes.
Finally in
\cite{gw4}
the general features of such echoes have been investigated in relation to quasi-normal modes. 

\textbullet\ Also imaging of the shadow of the black hole candidate at the center of the Milky Way will soon be possible with the Event Horizon Telescope, a tool that may already provide hints on the quantum nature of the near horizon geometry
\cite{eht}. 
Further to this, unexplained explosive phenomena have been observed in the Universe, for example, think at Fast Radio Bursts or ultra-luminous x-ray sources. At present there is no conclusive connection between any observed phenomena and the quantum gravity signature of exotic compact objects. However it is indeed possible that the experimental counterparts of these theoretical models have already been observed and not yet recognized.
Therefore these are exciting times for black holes physics, a better understanding of gravity in the strong field is emerging and will soon be tested experimentally, and it is indeed possible that, when enough data will be available, nature will surprise us with some unforeseen effect.


\end{document}